\documentclass[useAMS,usenatbib]{mn2e}

\usepackage{times}  % to get nice fonts for PDF
\usepackage{listings}
\usepackage{graphics}
\usepackage{subfigure}
\usepackage{graphicx}
\usepackage{amssymb}
\usepackage{hyperref}
\usepackage{aas_macros}

\def\gsim{\ga}
\def\lsim{\la}

\def\Msol{M_\odot}

\def\yr{{\rm yr}}

\def\crate{\dot M_{\rm cool}}

\def\simprop{ \lower .75ex \hbox{$\sim$} \llap{\raise .27ex \hbox{$\propto$}} }

\title[The impact of star formation laws]{
On the impact of empirical and theoretical star formation laws on galaxy formation} 
\author[Claudia del P. Lagos]{
\parbox[t]{\textwidth}{
\vspace{-1.0cm}
Claudia del P. Lagos$^{1}$,
Cedric G. Lacey$^{1}$,
Carlton M. Baugh$^{1}$,
Richard G. Bower$^{1}$,
Andrew J. Benson$^{2}$
}
\vspace*{6pt} \\
$^{1}$Institute for Computational Cosmology, Department of Physics, 
University of Durham, South Road, Durham, DH1 3LE, UK. \\
$^{2}$California Institute of Technology, Pasadena, CA 91125, USA.
\vspace*{-0.5cm}}

\begin{document}

%\date{Accepted ???. Received ???; in original form ???}

\pagerange{\pageref{firstpage}--\pageref{lastpage}} \pubyear{2010}

\maketitle

\label{firstpage}

\begin{abstract}
We investigate the consequences of applying different star formation
laws in the galaxy formation model {\texttt{GALFORM}}.  Three broad
star formation laws are implemented: the empirical relations of
Kennicutt and Schmidt and Blitz \& Rosolowsky and the theoretical
model of Krumholz, McKee \& Tumlinson. These laws have no free
parameters once calibrated against observations of the star formation
rate (SFR) and gas surface density in nearby galaxies.  We start from
published models, and investigate which observables are sensitive to a
change in the star formation law, without altering any other model
parameters.  {We show that changing the star formation law (i)
does not significantly affect either the star formation history of the
universe or the galaxy luminosity functions in the optical and
near-IR, due to an effective balance between the quiescent and burst
star formation modes; (ii) greatly affects the cold gas contents of
galaxies; (iii) changes the location of galaxies in the SFR versus
stellar mass plane, so that a second sequence of ``passive'' galaxies
arises, in addition to the known ``active'' sequence.  We show that
this plane can be used to discriminate between the star formation
laws.}
\end{abstract}

\section{Introduction}

A proper understanding of how galaxies form and evolve must include a
description of star formation and the physics that regulate
this phenomenon.  The lack of a theoretical
description of star formation (SF) has forced galaxy formation modellers to
adopt simple parametric recipes (e.g. \citealt{White91};
\citealt{Lacey93}; \citealt{Cole94}; \citealt{Kauffmann98};
\citealt{Cole00}; \citealt{Springel01}; \citealt{Cora06};
\citealt{Monaco07}; \citealt{Lagos08}; \citealt{Cattaneo08}).  

Two forms commonly adopted for the 
SF law relate the star formation rate
(SFR) per unit area, $\Sigma_{\rm SFR}$, to the gas surface density
either as: (i) $\Sigma_{\rm SFR} \propto \Sigma_{\rm gas}^N$
(\citealt{Schmidt59}), or (ii) $\Sigma_{\rm SFR} \propto \Sigma_{\rm
gas}/\tau_{\rm dyn}$ (\citealt{Shu73}), where $\tau_{\rm dyn}$ is a
dynamical timescale.
Observational constraints on the form of the SF law remained scarce for
many years. Early observations of individual galaxies gave
conflicting results for the power-law index $N$ in relation (i) above,
returning values over the range $N\approx1-3$ (see
\citealt{Sanduleak69} and \citealt{Hartwick71} for extreme examples).
Nor was it clear, when applying form (ii), whether a local or global
dynamical timescale $\tau_{\rm dyn}$ is more relevant (e.g. the local
free-fall time of the gas or the orbital time around a galaxy, see
\citealt{Madore77} and \citealt{Solomon88}, respectively).  The
situation improved with the observational sample constructed by
Kennicutt (1998, hereafter K98), which he used to examine correlations
between the global SFR and gas surface densities for a wide range of
nearby galaxies, from normal spirals to starbursts. K98 found that
the global relation was remarkably well fit by (i), with $N =
1.40 \pm 0.15$, but also nearly equally well fit by (ii),
when taking $\tau_{\rm dyn}$ equal to the disk orbital time.  
However, observational studies of the resolved radial profiles of
SFR and gas in individual galaxies showed a breakdown of the
power-law relation at low gas surface densities, with a sharp cut-off
in $\Sigma_{\rm SFR}$ below a critical gas surface density
\citep{Kennicutt89,Martin01}. This feature was interpreted using
simple gravitational arguments as marking the onset of
dynamical stability in the gas layer, thereby allowing the
fragmentation of the gas into star-forming clouds at gas surface
densities above a threshold value \citep{Toomre64}.

Over the past ten years, enormous advances in the characterisation 
of the SF law have been made possible through the
availability of new, high quality, spatially resolved observations of
HI \citep[e.g.][]{Walter08} and CO (e.g. \citealt{Helfer03};
\citealt{Leroy09}), and of UV and IR SFR tracers in samples 
of nearby galaxies. These data have
allowed the accurate estimation of the molecular and atomic hydrogen
content of galaxies over a wide range of morphologies and gas fractions.  
In addition, more reliable estimates of unobscured SF
in the UV \citep[e.g.][]{Gildepaz07} and of dust-obscured SF
in the IR \citep[e.g.][]{Calzetti07} have allowed better
determinations of the SFR, both globally and within
individual SF regions.  There is now support
for a correlation between $\Sigma_{\rm SFR}$ and the molecular gas
surface density, $\Sigma_{\rm mol}$, of the form $\Sigma_{\rm
SFR}\propto\Sigma_{\rm mol}$ (\citealt{Wong02}; 
\citealt{Kennicutt07}; \citealt{Bigiel08}). This relation is much
stronger than that with either the total cold gas 
or atomic hydrogen surface densities. 
At high redshifts, there are indications that the same
form, $\Sigma_{\rm SFR}\propto\Sigma_{\rm mol}$ could hold
(\citealt{Bouche07}; \citealt{Genzel10}).  The correlation of
$\Sigma_{\rm SFR}$ with $\Sigma_{\rm mol}$ seems physically reasonable
since stars are observed to form in dense molecular gas clouds (see
\citealt{Solomon05} for a review).

The threshold in the SF law suggested by the observations of 
\citet{Kennicutt89} and \citet{Martin01} has since been studied in 
more detail. SF activity has been observed in galaxies even at low
gas surface densities ($\Sigma_{\rm gas} \le 10 M_{\odot}\,\rm
pc^{-2}$; \citealt{Heyer04}; \citealt{Fumagalli08};
\citealt{Bigiel08}; \citealt{Roychowdhury09}; \citealt{Wyder09}), but
following a steeper relation in $\Sigma_{\rm gas}$ compared to the SF law found at higher
gas surface densities, suggesting a different regime of SF activity.
Furthermore, \citet{Blitz06} and \citet{Leroy08} found that the ratio
of molecular-to-atomic hydrogen surface densities, $R_{\rm mol}$, is
well fitted by a power-law function of the hydrostatic pressure in
galactic disks, $R_{\rm mol}\propto \rm P_{\rm ext}^{\alpha}$, with
the same relation holding from dwarf to normal spiral galaxies.  An
empirical SF law emerges from these studies in which the relationship
between $R_{\rm mol}$ and $P_{\rm ext}$ naturally divides the
$\Sigma_{\rm SFR}-\Sigma_{\rm gas}$ plane into two regimes of SF
activity when a linear relation such as $\Sigma_{\rm
SFR}\propto\Sigma_{\rm mol}$ is assumed, according to whether $R_{\rm
mol}>1$ or $<1$. In this picture, it is the transition from atomic- to
molecular-dominated gas which causes the change of slope in the
$\Sigma_{\rm SFR}-\Sigma_{\rm gas}$ relation, rather than the stability of the 
disk.  Note that the empirical SF laws discussed above
attempt to describe the SF activity in galaxies on scales larger than
$\sim 500\,\rm pc$, averaging over representative volumes of the
ISM including both star-forming clouds and intercloud gas
(e.g. \citealt{Onodera10}; \citealt{Fumagalli10};
\citealt{Schruba10}).

Even though the quality of the observational data has improved the
characterization of the SF law, there is still uncertainty regarding
the main physical mechanism that governs the SFR in galaxies.
Possibilities currently under discussion include the formation of
molecules on dust grains in turbulence-regulated giant molecular
clouds (GMCs; \citealt{Krumholz05}; \citealt{Krumholz09}) and 
collisions of GMCs in shearing disks (\citealt{Tan00};
\citealt{Schaye04}; \citealt{Silk09}; 
see \citealt{McKee07} for a comprehensive review of
the theory of SF activity). Direct comparisons of empirical and
theoretical SF models with observations by \citet{Leroy08} revealed
that thresholds of large scale stability, as proposed by
\citet{Kennicutt98}, or a simple dependence on the orbital or
free-fall timescales as in \citet{Tan00}, do not offer a good
explanation for the observed $\Sigma_{\rm SFR}\propto\Sigma_{\rm mol}$
relation.  On the other hand, for a suitable choice of parameters, the
\citet{Krumholz09} SF law is able to reproduce the observational
results described above fairly well. The nature of the physical
processes underlying the global SF law could be identified by
resolving individual GMCs and SF regions in the Milky Way and other
galaxies. Ongoing observations with Herschel and future telescopes
such as ALMA (Atacama Large Millimeter Array) and the SKA (Square
Kilometre Array), will help us to better characterise the properties
of individual SF regions, and to test the applicability of
locally-derived SF laws at higher redshifts.

In this paper, we implement new recipes for SF in the
{\texttt{GALFORM}} semi-analytical model of galaxy formation in a
$\Lambda$ Cold Dark Matter ($\Lambda$CDM) cosmology (\citealt{Cole00}).  
We focus on the empirical
laws proposed by \citet{Kennicutt98} and \citet{Blitz06} and the
theoretical model of \citet{Krumholz09}. Our aim is to identify the galaxy 
observable properties which can distinguish between different
SF prescriptions, and also to identify behaviour which is common to
different SF laws. To do this, we insert the new SF prescriptions into
existing \texttt{GALFORM} models \citep{Baugh05,Bower06} {\it without}
changing any other model parameters to achieve a new `best' fit to
observations (see \citealt{Cole00} and \citealt{Bower10} for
discussions of fitting model parameters). 
We first study predictions for the cosmic SFR evolution
and then for the evolution of the cold gas content of galaxies and 
their level of SF activity, {which are very sensitive to this choice. 
An important result obtained in this work is that 
the study of the SFR in galaxies of different stellar masses 
offers a useful constraint on the choice of SF law.} New
radio telescopes such as MeerKAT \citep{Booth09}, ASKAP
\citep{Johnston08} and SKA \citep{Schilizzi08} will measure the
abundance of HI in the Universe and its evolution. ALMA will soon
begin to probe both the molecular gas contents and dust-obscured SFRs
in high-redshift galaxies. The predictions made in this paper can be
used in conjunction with these future observations to constrain the
physical processes that trigger SF in galaxies.

This paper is organized as follows.  In Section~\ref{modelssec}, we
present the {\texttt{GALFORM}} model and the two variants we
use to study the impact of applying different SF
laws, pointing out the main differences and similarities between them.
We also describe the SF laws tested.
In Section~\ref{SFRevoSec} we study the impact of
changing the SF law on the evolution of the cosmic SFR density.  
Section~\ref{coldgascontent} examines the evolution of the cold gas 
content of galaxies, and we discuss the physics that shapes its evolution. 
In Section~\ref{SFHgals}, {we present a new means to distinguish between 
different SF laws, namely the shape of the sequences in the $\rm SFR$-stellar
mass plane,} and compare with available observations. Finally, in
Section~\ref{discussion}, we discuss our results in the context of
future surveys and observations. The appendices cover the details of 
the numerical scheme used to implement the new SF laws, some details 
of the SF laws used and present 
some predictions for observables which are insensitive to the choice 
of SF law.

\section{Modelling the star formation activity in galaxies}
\label{modelssec}

We first give an overview of the {\texttt{GALFORM}} 
galaxy formation model (\S~2.1), 
before describing the star formation recipes used 
in {\texttt{GALFORM}} (\S~2.2). In \S~2.3 we outline the new star 
formation laws {(see Appendix~\ref{numer:integration} for 
a description of the numerical implementation of these laws and Appendix~\ref{App:sfls} 
for some more details of the SF laws).}

\subsection{The galaxy formation model}
\label{sec:galform}

We study the impact on the galaxy population of applying different
star formation laws in the {\texttt{GALFORM}} semi-analytic model of
galaxy formation (\citealt{Cole00}; \citealt{Benson03};
\citealt{Baugh05}; \citealt{Bower06}; \citealt{Lacey08}; \citealt{Benson10}). 
{\texttt{GALFORM}} models the main physical
processes that shape the formation and evolution of galaxies: 
(i) the collapse and merging of dark matter (DM) halos, (ii) the
shock-heating and radiative cooling of gas inside DM halos, leading to
the formation of galactic disks, (iii) quiescent star formation in
galactic disks, (iv) feedback from supernovae (SNe), from AGN heating
and from photo-ionization of the IGM, (v) chemical enrichment of stars
and gas, and (vi) galaxy mergers driven by dynamical friction within
common DM halos, which can lead to the formation of spheroids and 
trigger bursts of SF (for reviews see
\citealt{Baugh06}; \citealt{Benson10b}).  Galaxy luminosities are
computed from the predicted star formation and chemical enrichment
histories using a stellar population synthesis model.  Dust extinction
is calculated self-consistently from the gas and metal contents of 
each galaxy and the predicted scale lengths of the disk and bulge 
components using a radiative transfer model (see \citealt{Cole00} and \citealt{Lacey10b}).

We consider the models published by Baugh et al. (2005; hereafter Bau05) 
and Bower et al. (2006, hereafter Bow06).  The main successes of the 
Bow06 model are the reproduction of the observed break in
the galaxy luminosity function (LF) at $z=0$ in the $b_{J}$- and $K$-bands,
the observational inferred evolution of the 
stellar mass functions up to $z\approx 5$, the bimodality in the
colour-magnitude relation and the abundance and properties of red
galaxies (\citealt{Gonzalez-Perez09}).  The Bau05
model likewise matches $b_{J}$- and $K$-band LFs at $z=0$, but also
the number counts and redshift distribution of sub-millimetre
selected galaxies (see also \citealt{Gonzalez10}), disk sizes
(\citealt{Almeida07}; \citealt{Gonzalez09}), the evolution of
Ly$\alpha$-emitters (\citealt{Delliou06}; \citealt{Orsi08}), counts,
redshift distributions and LFs of galaxies selected in the mid- and
far-IR \citep{Lacey08,Lacey10a} and the Lyman-break galaxy LF at
redshifts $z=3-10$ \citep{Lacey10b}.

We refer the reader to the original papers for full details of the
Bau05 and Bow06 models. Discussions of the differences
between the models can be found in \citet{Almeida07},
\citet{Gonzalez09}, \citet{Gonzalez-Perez09} and \citet{Lacey10b}.
Briefly, the main differences between the models are:
\begin{itemize} 
\item The SF timescale in galactic disks.  The Bow06 model adopts a
parametrization for the SF timescale which depends on the global
dynamical timescale of the disk, whereas the Bau05 model adopts a
timescale which depends only on the disk circular velocity (see $\S 2.1$).  
Consequently, high redshift galaxies in the Bau05 model have longer 
SF timescales than those in the Bow06 model and are therefore 
more gas-rich.  This leads to more
SF activity in bursts in the Bau05 model compared to the Bow06 model.
\item The stellar initial mass function (IMF) is assumed to be universal in
the Bow06 model, with the \citet{Kennicutt83} IMF adopted. The Bau05 model 
assumes two IMFs, a top-heavy IMF during starbursts (SB) and
a Kennicutt IMF during quiescent SF in disks. Defining the IMF slope
$x$ through $dN/d\ln m \propto m^{-x}$, the Kennicutt IMF has $x=0.4$
below $m= 1 \Msol$ and $x=1.5$ at higher masses, 
while the top-heavy IMF has $x=0$. 
\item The gas expelled from galaxies by SN feedback is assumed to be 
reincorporated into the hot halo after a few dynamical times in the Bow06 
model.  In the Bau05 model, this happens only after the DM
halo doubles its mass.  This time, called the halo lifetime, is
usually much longer than the dynamical time of the halo.
\item In order to reproduce the bright-end of the LF, the Bau05 model
invokes SNe driven superwinds with a mass ejection rate proportional
to the SFR (note that the expelled material is not reincorporated into
any subsequent halo in the DM merger tree).  The Bow06 model includes
AGN feedback to switch off the cooling flow in haloes
where the central black hole (BH) has an Eddington luminosity
exceeding a multiple of the cooling luminosity (the multiple being a
model parameter).
\item The Bow06 model includes disk instabilities
(e.g. \citealt{Mo98}) as an extra mechanism to trigger bursts of SF
and to make spheroids, and hence to build BH mass \citep{Fanidakis10}.
\end{itemize}

We use halo merger trees extracted from the Millennium cosmological N-body 
simulation \citep{Springel05} in the Bow06 model and Monte-Carlo trees in the
Bau05 model, as in the original models. We use the Monte-Carlo algorithm of
\cite{Parkinson08} to generate DM merger trees for the Bau05 model 
from that of \citet{Cole00}. \cite{Gonzalez10}
tested that changing the tree-generation scheme does not significantly
change the model predictions.
The Bau05 model adopts a $\Lambda$CDM cosmology with a present-day matter 
density parameter,
$\Omega_{\rm m} = 0.3$, a cosmological constant,
$\Omega_{\Lambda}=0.7$, a baryon density, $\Omega_{\rm baryons}=
0.04$, a Hubble constant $h = 0.7$, and a power spectrum normalization
of $\sigma_{8}= 0.93$.  The Bow06 model is set in the Millennium
simulation which has the following cosmology:
$\Omega_{\rm m}=\Omega_{\rm DM}+\Omega_{\rm baryons}=0.25$ (giving a
baryon fraction of $0.18$), $\Omega_{\Lambda}=0.75$, $\sigma_{8}=0.9$
and $h=0.73$.  The resolution of the $N$-body
simulation is fixed at a halo mass of $1.72 \times 10^{10} h^{-1} M_{\odot}$. 
In the Bau05 model, Monte-Carlo merger histories are generated for a grid of halo
masses. The grid is laid down at selected redshifts, sampling
haloes in a fixed dynamic range of mass which scales as $(1+z)^{-3}$
to approximately track the break in the mass function. The
minimum halo mass for merger trees at $z=0$ is $10^{9} h^{-1}
M_{\odot}$. The Monte-Carlo trees therefore include lower mass haloes
than in the N-body case.

\subsection{The original star formation laws in {\texttt{GALFORM}}}

The original {\texttt{GALFORM}} models used 
a parametric SF law to compute quiescent SFR at each timestep \citep{Cole00} 

\begin{equation}
\psi=\frac{M_{\rm cold}}{\tau_{\star}},
\label{SFlawGALFORM}
\end{equation}

\noindent where $M_{\rm cold}$ is the total cold gas mass in the
galactic disk and $\tau_{\star}$ is the timescale for the SF
activity which may depend on galaxy parameters but not on the gas mass. 
This choice was motivated by the observations of
K98 (see \citealt{Bell03}). The SF timescale in  
Eq.~\ref{SFlawGALFORM} was parametrized by \citet{Cole00} as,

\begin{equation}
\tau_{\star}= \frac{\tau_{\rm disk}}{\epsilon_{\star}} \, (V_{\rm
  disk}/V_{0})^{\alpha_{\star}}, 
\label{SFRold2}
\end{equation}

\noindent where $V_{0}=200 \,\rm km\, s^{-1}$, $V_{\rm disk}$ is the
disk circular velocity at the half mass radius, $\tau_{\rm
disk}=r_{\rm disk}/V_{\rm disk}$ is the disk dynamical timescale, and
$\epsilon_{\star}$ and $\alpha_{\star}$ are free parameters, with
values originally chosen to reproduce the gas-to-luminosity ratio 
observed in local spiral galaxies \citep{Cole00}.  Bow06 adopted values of 
$\epsilon_{\star}=0.0029$ and $\alpha_{\star}=-1.5$, mainly to reproduce 
the galaxy LF, without particular reference to the cold gas to 
luminosity ratio (see \citealt{Kim10}).

In the Bau05 model a different SF timescale was adopted, in order to
reduce the amount of quiescent SF activity at high redshift, thereby
allowing more SF in bursts.  The recipe used in the Bau05 model is

\begin{equation}
\tau_{\star}=\tau_{0}\, (V_{\rm disk}/V_{0})^{\alpha_{\star}},
\label{SFRold}
\end{equation}

\noindent where $V_{0}=200 \,\rm km\, s^{-1}$, $\tau_{0}$ and $\alpha$
are free parameters set to $\tau_{0}=8 \, \rm Gyr$ and
$\alpha_{\star}=-3$, in order to match the local gas
mass to luminosity ratio (see \citealt{Power09}).

During SBs caused by galaxy mergers and, in the case of the Bow06
model, also by disk instabilities, the available
reservoir of cold gas is assumed to be consumed in the SB event with a finite
duration. The SF timescale in bursts is taken to be
$f_{\rm dyn} \tau_{\rm bulge}$, with a floor value $\tau_{\rm *
burst,min}$, with $\tau_{\rm bulge}$ being the bulge dynamical
time. Bau05 adopted $f_{\rm dyn} = 50$ and $\tau_{\rm * burst,min}=0.2$~Gyr 
while Bow06 used $f_{\rm dyn} = 2$ and $\tau_{\rm * burst,min}=0.005$~Gyr.

The need for two free parameters in Eqs.~\ref{SFRold2} and \ref{SFRold} 
reflects a lack of understanding of the physics of SF. New, high resolution, 
spatially resolved data reveal that a SF law of the form
of Eq.~\ref{SFlawGALFORM} with $\tau_{\star}$ independent of the gas density 
results in a poor fit to the observations
(\citealt{Leroy08}). It is necessary to revisit the SF law
used in {\texttt{GALFORM}} and to study more consistent ways to
characterise the SF in galaxies.

\subsection{The new star formation laws}
\label{models}

Here we summarise the three new forms of SF law that we
implement for quiescent SF 
in {\texttt{GALFORM}}. These are the empirical relations of
(i) \citet{Kennicutt98} and (ii) \citet{Blitz06}, and (iii) the
theoretical model of \citet{Krumholz09}.  
The observational situation for starbursts is less clear than it is
for disks, so we retain the original \texttt{GALFORM} prescription for
the star formation timescale in SB (see above).

\subsubsection{The Kennicutt-Schmidt law}

Following the pioneering work of \citet{Schmidt59}, observational
studies have found that the surface density of SF ($\Sigma_{\rm SFR}$)
correlates with the projected gas density ($\Sigma_{\rm gas}$). 
K98 fitted the relation 

\begin{eqnarray}
\Sigma_{\rm SFR}=A \,\Sigma^N_{\rm gas},
\label{sch}
\end{eqnarray}
where $N=1.4 \pm 0.15$ and $A=0.147$\footnote{We have rescaled
the value of $A$ to account for the difference between the
Salpeter IMF assumed by K98 and the Kennicutt IMF used in our
model for quiescent SF. The SFRs used in K98 were inferred
from H$\alpha$ luminosities, and for a fixed H$\alpha$ luminosity, the SFR on
assuming a Kennicutt IMF is $0.91$ times that inferred on adopting a
Salpeter IMF. Note also that we always use gas surface densities
including helium, assuming a hydrogen mass fraction of 74\%.} 
when $\Sigma_{\rm gas}$ and
$\Sigma_{\rm SFR}$ are measured in $M_{\odot}\rm pc^{-2}$ and
$M_{\odot}\, \rm pc^{-2} Gyr^{-1}$, respectively. 
This
relation holds over $5$ orders of magnitude in SFR and gas surface
density, but show a break to a steeper relation in the outer regions of spirals 
and in
dwarf galaxies (K89, K98, \citealt{Bigiel08,Leroy08}). 

The Kennicutt-Schmidt SF law combines the power-law dependence of the
SFR on $\Sigma_{\rm gas}$ at high gas densities with a cut-off in SF
below a critical gas surface density, $\Sigma_{\rm crit}$, as observed
at low gas surface densities by K89 and \citet{Martin01}. The
$\Sigma_{\rm crit}$ threshold is motivated by the \cite{Toomre64}
stability criterion. In the case of a thin isothermal gas disk 
{with a flat rotation curve}, the
critical surface density for gravitational instability of axisymmetric
perturbations is given by

\begin{equation}
\Sigma_{\rm crit}= \frac{\sqrt{2}}{Q_{\rm crit} \,\pi G} \, \sigma_{\rm g}\, \frac{V}{R},
\label{gascut}
\end{equation}

\noindent {where $\rm Q_{\rm crit}$ is a dimensionless constant $\sim 1$ and 
$\sigma_{\rm g}$ is the velocity dispersion of the gas (see Appendix~\ref{App:K98} 
for a derivation of Eq.~\ref{gascut}).} We adopt 
$\sigma_{\rm g}=10 \, \rm km\, 
\rm s^{-1}$ consistent with the observations of \citet{Leroy08}. 
{K98 found $Q_{\rm crit}\approx 2.5$ (after
scaling by our choice of $\sigma_{\rm g}$, which is larger than the
original value adopted by K98). By adopting these values, we calculate $\Sigma_{\rm crit}$,  
and suppress SF at radii in which $\Sigma_{\rm gas}<\Sigma_{\rm crit}$.}

We will consider two forms 
of the Kennicutt-Schmidt law, with and without SF suppression 
below $\Sigma_{\rm crit}$ (Eq.~\ref{gascut}). 
We denote these SF laws as KS.thresh (with
suppression) and KS (without suppression).

\subsubsection{The Blitz \& Rosolowsky model: molecular gas 
fraction determined by pressure}

Blitz \& Rosolowsky (2006, hereafter BR06) based their SF law on two
observationally motivated considerations: (i) Observations in the IR
and at millimetre wavelengths suggest that stars form in dense gas
environments, namely GMCs (see \citealt{Solomon05} for a review). BR06
assume that the SFR is set by the surface density of molecular
gas\footnote{Note that we include the associated helium in the
molecular gas mass, in the same way as for the total gas mass.}
$\Sigma_{\rm mol}$, with a proportionality factor (given as an inverse
timescale) $\nu_{\rm SF}$:
\begin{equation}
\Sigma_{\rm SFR} = \nu_{\rm SF}\, \Sigma_{\rm mol}.
\label{sfrblitz}
\end{equation}
(ii) BR06 propose that the ratio of molecular to atomic hydrogen gas,
$R_{\rm mol}$, is given by a power law in the internal hydrostatic
pressure in a galactic disk, $P_{\rm ext}$,
\begin{equation}
R_{\rm mol} \equiv \frac{\Sigma(\rm H_2)}{\Sigma(\rm HI)} = \left( \frac{P_{\rm
ext}}{P_{0}}\right)^{\alpha}.
\label{fmol}
\end{equation}
BR06 found that the observed molecular-to-atomic ratios in their
galaxy sample were well fitted using values of $\rm log(P_{0}/k_{\rm B}
[\rm cm^{-3}\rm K])=4.54\pm 0.07$ and $\alpha=0.92\pm
0.07$. \citet{Leroy08} found similar values using a somewhat larger
sample. The hydrostatic pressure in disk galaxies at the
mid-plane is calculated following Elmegreen (1993; {see Appendix~\ref{App:BR} 
for details}).

The original SF law of Eq.~\ref{sfrblitz} can hence be rewritten in
terms of the total gas surface density and the molecular-to-total
hydrogen ratio, $f_{\rm mol}= \Sigma_{\rm mol}/\Sigma_{\rm gas} =
R_{\rm mol}/(R_{\rm mol}+1)$, as

\begin{equation}
\Sigma_{\rm SFR} = \nu_{\rm SF} \, f_{\rm mol}\, \Sigma_{\rm gas}.
\label{SFlawBR}
\end{equation}
We consider two cases for $\nu_{\rm SF}$; (i) a constant value
$\nu_{\rm SF}= \nu^{0}_{\rm SF}=0.525 \pm 0.25\,\rm Gyr^{-1}$
\citep{Leroy08}\footnote{\citet{Leroy08} derived their SFRs from a
calibration using far-UV luminosities, assuming a \citet{Kroupa01}
IMF. For the Kennicutt IMF, we infer SFRs 1.02 times larger. For
simplicity, we do not apply this correction factor, since it is so
close to 1.}, which gives a linear dependence of $\Sigma_{\rm SFR}$ on
$\Sigma_{\rm gas}$ at high surface densities; (ii) a surface density
dependent $\nu_{\rm SF}$ given by
\begin{equation} 
\nu_{\rm SF}=\nu^{0}_{\rm SF}\, \left[1+\left(\frac{\Sigma_{\rm gas}}{\Sigma_{0}}\right)^{\rm q}\right],
\label{fitD09}
\end{equation}

\noindent where $\Sigma_{0}=200\,\rm M_{\odot}\,\rm pc^{-2}$ and $\rm
q=0.4$ are chosen to recover the K98 law at high gas surface
densities, and the steepening seen in the $\Sigma_{\rm SFR}$ --
$\Sigma_{\rm gas}$ relation when starbursts are included
\citep{Bigiel08}. The surface density $\Sigma_{0}$ is similar to the
typical surface densities of individual GMCs in spiral galaxies, so the
transition to the steeper dependence could be interpreted as happening
when these clouds start to overlap.  We refer to the SF law with a
constant $\nu_{\rm SF}$ as BR and to the second form, where $\nu_{\rm
SF}=\nu_{\rm SF}(\Sigma_{\rm gas})$, as BR.nonlin.

\subsubsection{The Krumholz, McKee \& Tumlinson model:
  turbulence-regulated star formation activity} 

Krumholz, McKee \& Tumlinson (2009; hereafter KMT09) 
calculate $\nu_{\rm SF}$ and $f_{\rm mol}$ of Eq~\ref{SFlawBR} for 
{a spherical cloud with SF regulated by supersonic turbulence.}

KMT09 assume $f_{\rm mol}$ is determined by the balance between the
photo-dissociation of $H_2$ molecules by the interstellar far-UV
radiation and the formation of molecules on the surface of dust
grains. \citet{Krumholz09b} calculated $f_{\rm mol}$ theoretically for
spherical cloud, and showed that it is approximately a function of the
gas surface density of the atomic-molecular complex (or GMC)
$\Sigma_{\rm comp}$ and of the gas metallicity $Z$ (see also
\citealt{McKee10}).  We use eqn.(2) from KMT09 for $f_{\rm
mol}(\Sigma_{\rm comp}, Z)$. KMT09 assume that the surface density of
the GMC is related to that of the ISM on larger scales by $\Sigma_{\rm
comp} = c \Sigma_{\rm gas}$, where the clumping factor $c \ge 1$ is a
free parameter, which is not predicted by the model.  It is implicitly
assumed that the fraction of the ISM in GMCs is equal to the molecular
fraction within a single GMC.  {Given that the KMT09 formula
predicts that $f_{\rm mol} \to 0$ in very metal poor environments, we
assume that a minimum $f_{\rm mol}^{\rm min} = 10^{-4}$ characterises
pristine gas at very high redshift (see Appendix~\ref{App:KMT1} for
more details).}

{KMT09 define $\nu_{\rm SF}$ as the inverse of the timescale required to convert 
all of the gas in a cloud into stars, and obtain it  
from a theoretical model of turbulent
fragmentation \citep{Krumholz05}. 
The parameter $\nu_{\rm SF}$ 
in a cloud 
depends on the 
cloud surface density, $\Sigma_{cl}$. KMT09 assume that 
$\Sigma_{cl}$ is a constant ($\Sigma_0$) in normal spiral galaxies, 
but increases in higher density environments, so that $\Sigma_{cl} = 
\max[\Sigma_0,\Sigma_{\rm gas}]$ (see Appendix~\ref{App:KMT2} for 
more details).} Combining all of these ingredients, KMT09 obtain: 
\begin{eqnarray}
% Second equation
\nu_{\rm SF}(\Sigma_{\rm gas}) & = & 
\nu^{0}_{\rm SF}\times 
\left(\frac{\Sigma_{\rm gas}}{\Sigma_{\rm 0}}\right)^{-0.33} \quad {\rm for } \quad \Sigma_{\rm gas} < \Sigma_{\rm 0} \\
& = & \nu^{0}_{\rm SF}\times \left(\frac{\Sigma_{\rm gas}}{\Sigma_{\rm
    0}}\right)^{0.33}  \quad \quad {\rm for} \quad \Sigma_{\rm gas} > \Sigma_{\rm 0} 
\label{sfreq2}
\end{eqnarray}
with $\nu^{0}_{\rm SF}= 0.38\,\rm Gyr^{-1}$ and $\Sigma_{\rm
0}=85\,M_{\odot}\,\rm pc^{-2}$. The molecular fraction $f_{\rm
mol}(c\Sigma_{\rm gas}, Z)$ contains the dependence on the clumping
factor $c$. The KMT09 SF model thus predicts three regimes of SF
(qualitatively similar to the BR.nonlin model): (i) regions with low
gas surface densities and hence low molecular gas fractions, which
produce steeper SF relations; (ii) an intermediate density regime
characterised by constant GMC surface densities; (iii) a high density regime
where there is an increase in the GMC density.  This model is able to
reproduce reasonably well the observed trend of $\Sigma_{\rm SFR}$
with $\Sigma_{\rm gas}$ in nearby galaxies on adopting $c\approx 5$, roughly consistent
with observed overdensities of molecular gas complexes
(\citealt{Rosolowsky05}; \citealt{Schuster07}). Hereafter we will
refer to this SF law as KMT.
 
\section{The evolution of the SFR density}
\label{SFRevoSec}
\begin{figure*}
\begin{center}
\includegraphics[trim = 3mm 3mm 1.5mm 1.5mm,clip,width=0.38\textwidth]{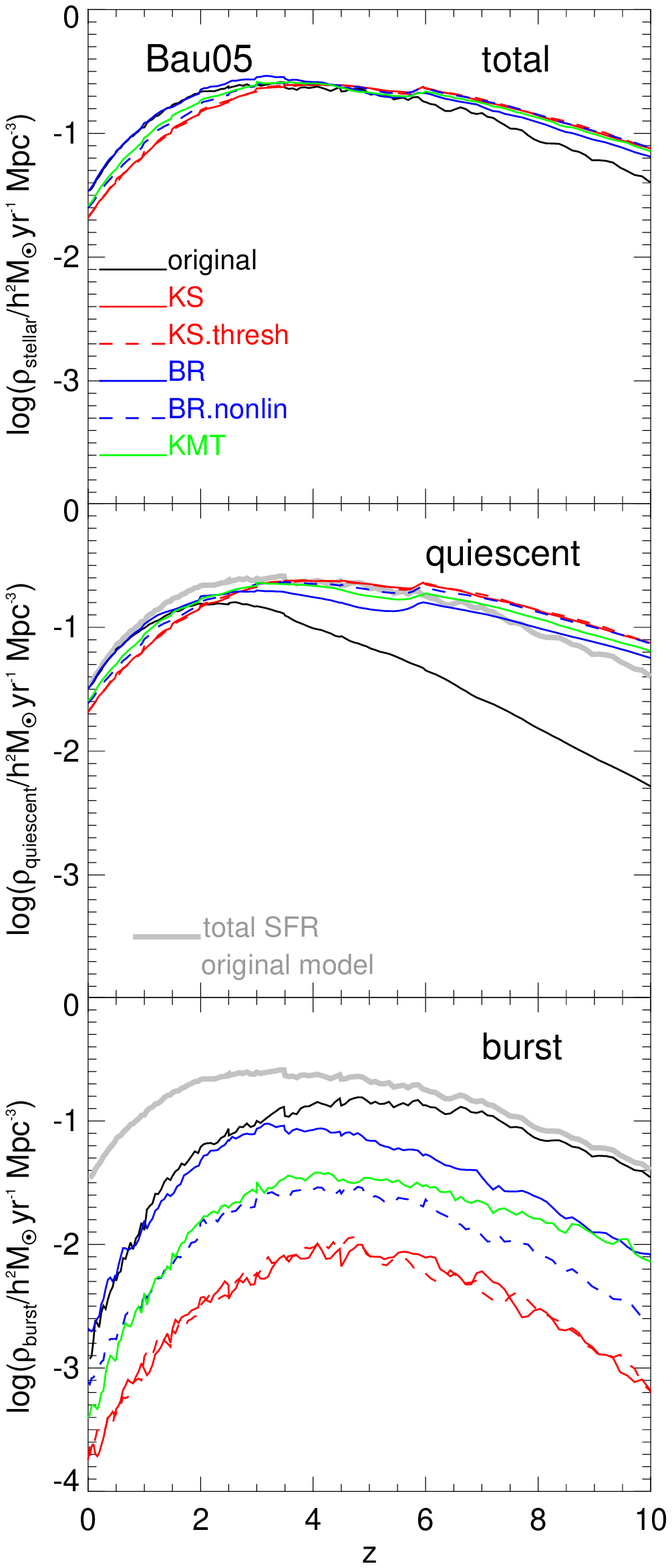}
\includegraphics[trim = 3mm 3mm 1.5mm 1.5mm,clip,width=0.38\textwidth]{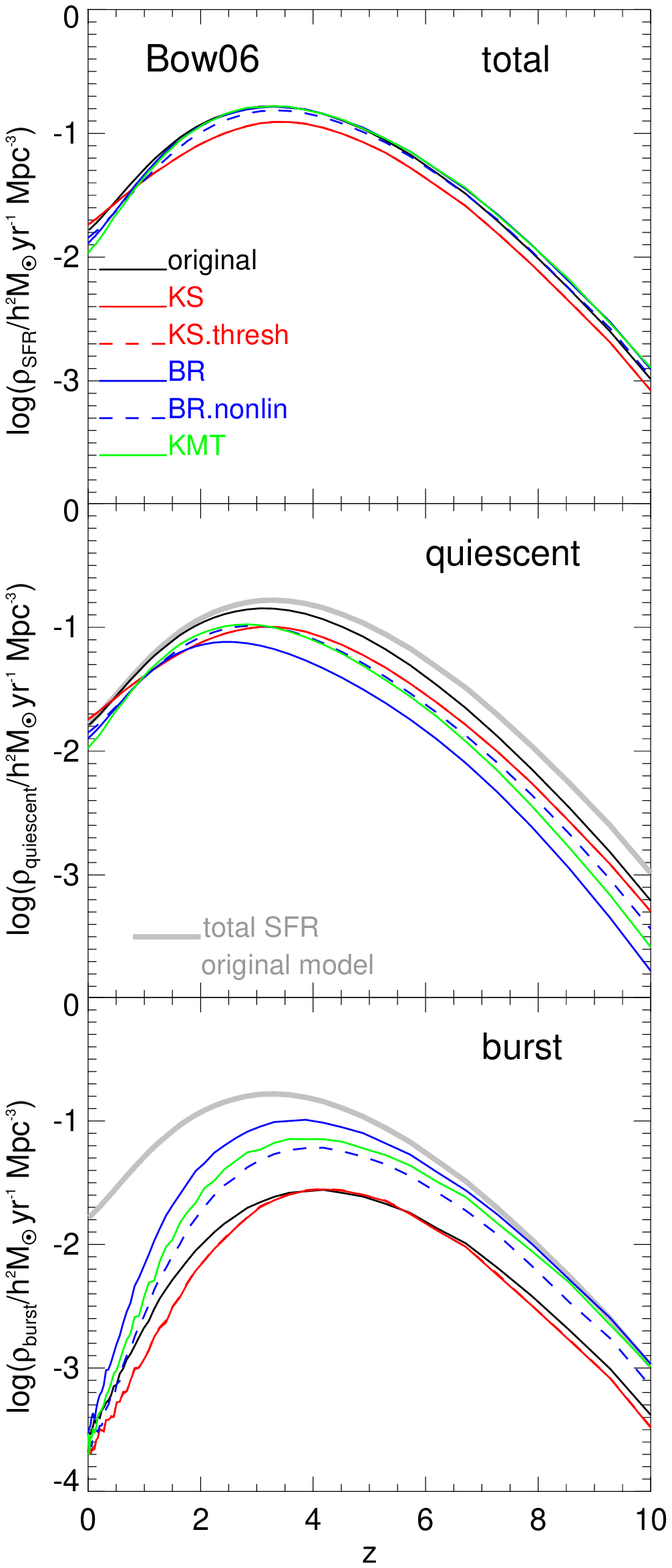}
\caption{ The evolution of the cosmic star formation
rate per unit volume. The left-hand panels show the Bau05 model and
its variants, and the right-hand panels show the Bow06 model. The top
row shows the total star formation rate, the middle panels show the
quiescent star formation and the bottom panels the star formation in
bursts.  For reference, in the middle and bottom panels the total SFR
density in the original models is shown by a thick gray solid line.}
\label{SFRevo}
\end{center}
\end{figure*}

Our aim is to investigate the impact on galaxy properties 
of changing the SF law 
over a wide range of redshifts, and to compare the new predictions 
with those of the original Bau05 and Bow06 models. To do this, we leave the
parameter values for processes other than the SF law unchanged, to isolate 
the effects of changing the SF prescription.  We apply the new SF laws consistently 
throughout the galaxy formation calculation {(see Appendix~\ref{App:SFR} for 
an illustration of applying the SF laws to the original models over a single 
timestep)}. 

Fig.~\ref{SFRevo} shows the evolution of the cosmic SFR density (top), 
distinguishing between quiescent (middle) and burst
(bottom) SF modes, for the Bau05 (left) and the Bow06
(right) models.  In the Bau05 model, the peak in SF activity is
at $z\approx2.5$. The peak location (though not its
height) is somewhat sensitive to the choice of SF law, moving to
higher redshift in the variant models.  The extreme cases are the KS
and KS.thresh SF laws, in which the peak shifts to $z\approx3.5$. In
the case of the KMT and BR.nonlin laws, the peak moves slightly to
$z\approx2.8$, while the BR law gives a very similar SFR density
evolution to the original Bau05 model.  The change in the location of
the SFR density peak is driven by the change in the level of quiescent
SF activity.  SF laws that produce higher SFR densities in the
quiescent mode at high redshift (i.e. KS, KS.thresh, KMT, BR.nonlin)
are characterised by a SFR-gas density correlation $\Sigma_{\rm
SFR}-\Sigma_{\rm gas}$ steeper than linear (e.g. $N>1$ in
Eq.~\ref{sch}). A consequence of the larger quiescent SFR at high
redshifts is less cold gas available to fuel starbursts,
and so the SFR density in the burst mode drops.

In the Bow06 model (right panels in Fig.~\ref{SFRevo}),
the evolution of the total SFR density is largely unaffected by the
choice of SF law even though the relative importance of the quiescent
and burst SF modes changes by more than one order of magnitude.  This
is due to the inclusion of starbursts triggered by disk instabilities ($\S 2$).  
The smaller SFR density in the quiescent mode relative to the original model when using 
the BR, BR.nonlin and KMT SF laws (middle panel),
leads to significant amounts of cold gas in galaxy disks.
This results in a higher frequency of unstable disks and gas rich
mergers, and also more prominent SBs due to the larger gas reservoirs.  
This increase in starburst activity
compensates for the lower quiescent activity.  In the case of the KS
and KS.thresh laws, the slightly smaller SFR density (by $\approx
-0.1$~dex) in the quiescent mode is not offset by an increase in the
SF activity from disk instabilities. This is because the
condition for a disk to become dynamically unstable is only weakly
dependent on mass ($\propto M_{\rm disk}^{-1/2}$). Hence a
substantial change in disk mass is needed to make a galaxy unstable,
and on average the change in disk mass is less than $0.1$~dex for the
KS and the KS.thresh laws, while for the other SF laws it is at
least $0.4$~dex.  Remarkably, all of the new SF laws produce peaks of
burst and quiescent SF activity separated by at least $\Delta z\approx
1$, while in the original Bow06 model, the peaks overlap.

The differences on applying the new SF laws in the Bau05 and
Bow06 models indicate the role of physical processes other than
quiescent SF in shaping the evolution of galaxies, e.g. disk
instabilities, and support our decision to leave the other model 
parameters unchanged. However, there are remarkable similarities between the
predictions which result on using different SF laws. Even though the
total density of SF activity is unchanged, the relative
importance of the burst and quiescent modes depends sensitively on the
choice of SF law, due mainly to the change in the cold gas content of
galaxies (see $\S 4$). The insensitivity of the total SFR density to the choice 
of SF law has been noted in previous work
(e.g. \citealt{Stringer07}; \citealt{Dutton09};
\citealt{Schaye10}). This can be understood as resulting
from the SFR (combined with the rate of gas ejection from galaxies due
to SN feedback, which is proportional to the SFR) on average adjusting
to balance the rate of accretion of cooling gas.
The SFR will tend to adjust in this way whenever the gas
consumption timescale $\tau_{\rm SF} \equiv M_{\rm cold}/SFR$ is less
than the age of the universe at that redshift. 
{A direct consequence of this interplay between the modes of SF 
is the insensitivity of the optical and near-IR galaxy LF and optical colours 
to the choice of SF law (see Appendix~\ref{localprops} 
for a comparison of the predicted LFs and other galaxy properties when using 
different SF laws).}

\section{Cold gas mass content of galaxies}
\label{coldgascontent}

The cold gas mass content of galaxies is sensitive to the choice of SF
law, since this determines the rate at which gas 
is converted into stars. Here we focus on three separate
observational probes of gas: the cold gas mass
function (CMF), the evolution of the cold gas density of the universe
and the gas-to-luminosity ratio in galaxies.

\subsection{Cold gas mass function}

\begin{figure}
\begin{center}
\includegraphics[trim = 3mm 1mm 1.5mm 1.5mm,clip,width=0.45\textwidth]{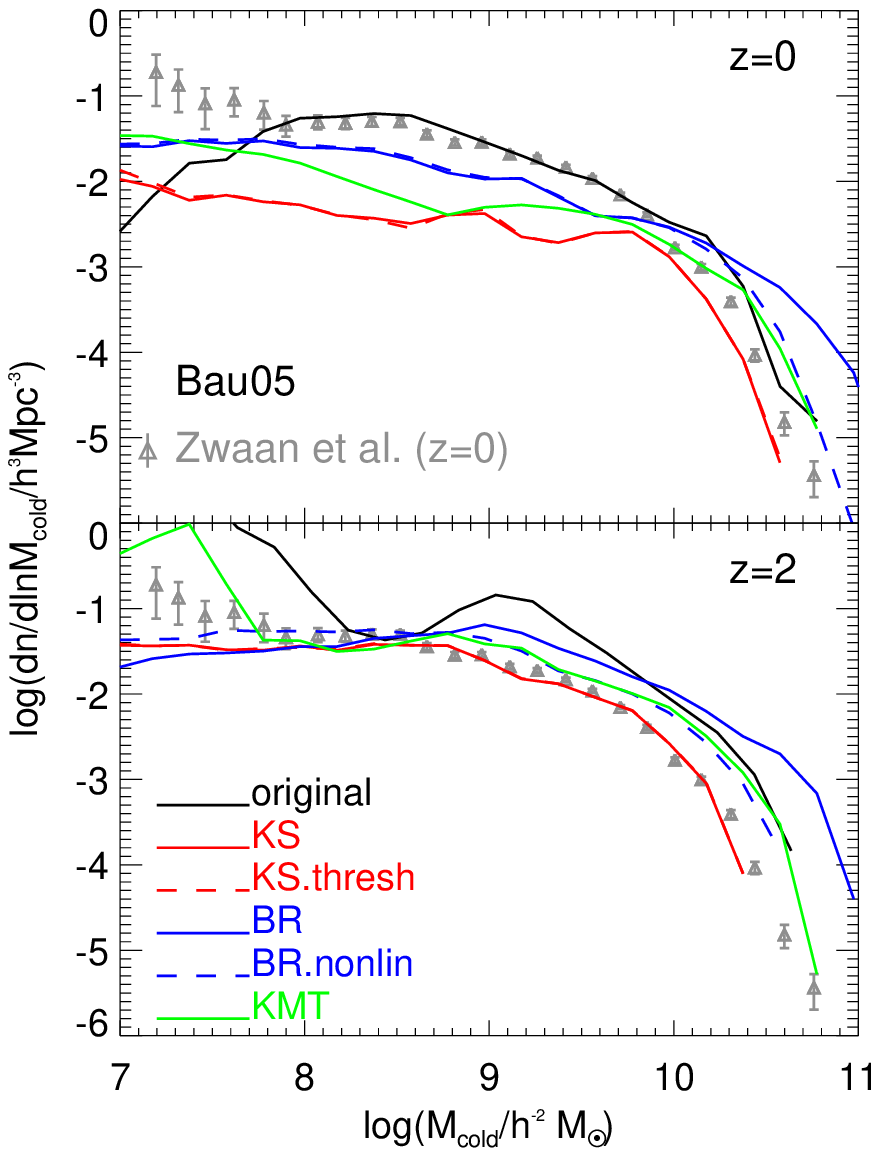}
\includegraphics[trim = 3mm 1mm 1.5mm 1.5mm,clip,width=0.45\textwidth]{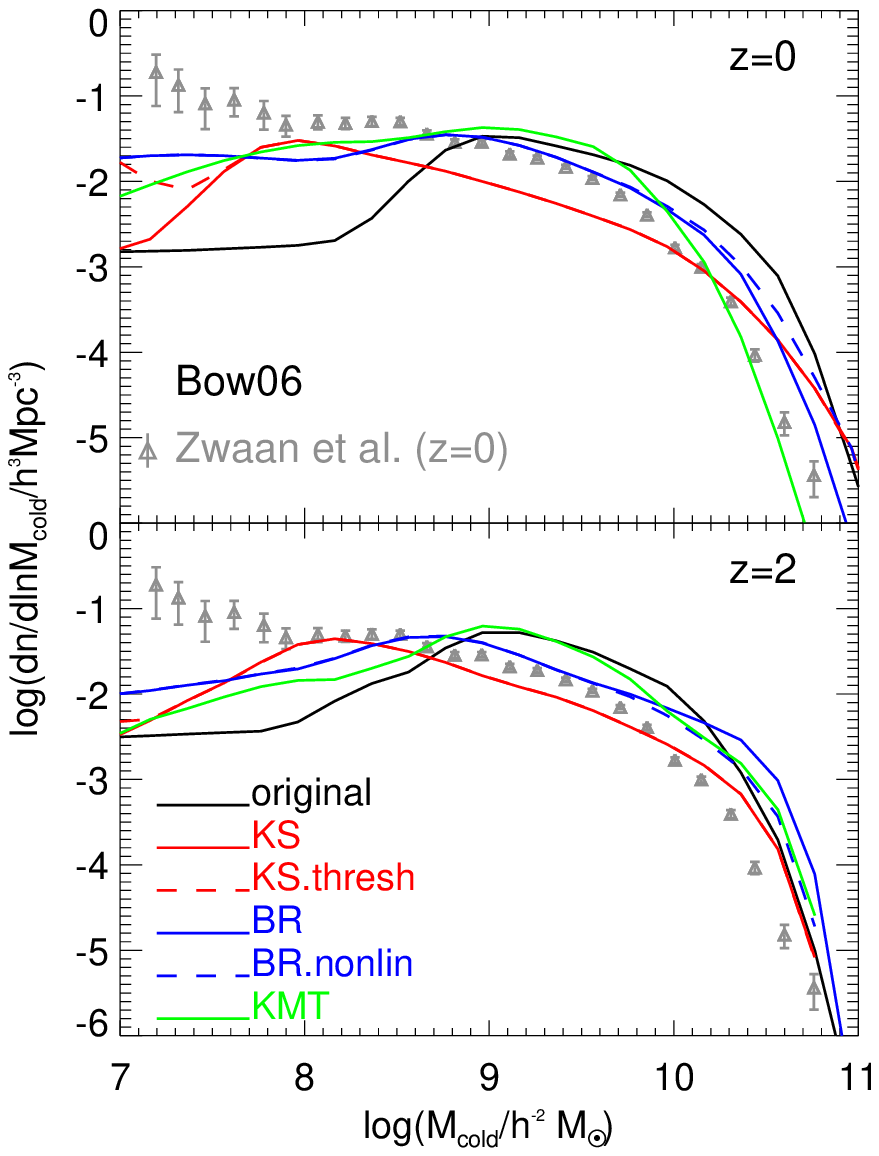}
\caption{The cold gas mass functions at $z=0$ and $z=2$ for the Bau05
(upper two panels) and the Bow06 (lower two panels) models and for the
different forms of the SF law. The line colours and styles are as
in Fig.~\ref{SFRevo}.  For reference, the observational estimate of the $z=0$ 
CMF from \citet{Zwaan05} is shown using grey symbols in all
panels. A constant $\rm H_2/HI$ ratio of 0.4 has been assumed to
convert the HI observations into cold gas masses.}
\label{CMFevo}
\end{center}
\end{figure}

Fig.~\ref{CMFevo} shows the CMF at different redshifts for the
original models and with different SF laws.  
The observed $z=0$ CMF of \citet{Zwaan05} is plotted to highlight 
evolution. Note that the observed
CMF plotted here is derived from the observed HI mass function assuming a
constant $\rm H_2/\rm HI$ ratio of $0.4$ and a hydrogen mass fraction
$X=0.74$ ($\S 2.3.1$) to convert the HI measurements into cold gas
masses (\citealt{Baugh04}; \citealt{Power09}; \citealt{Kim10}).

The Bau05 model is in
quite good agreement with the observed CMF, except at the lowest gas
masses. The parameters in the original model were
chosen to match the gas-to-luminosity
ratios discussed in \S\ref{sec:gasfrac}. The modified SF laws all lead
to poorer agreement with observations. 
On the other hand, the original Bow06 model 
is in poor agreement with the observed CMF. Some of the variant SF 
laws in this case lead to better agreement with the observed CMF.

Most of the changes in the CMF with the new SF laws can be
understood as resulting from the changed SF timescales in
disks, defined as $\tau_{SF} = M_{\rm cold}/SFR$. (Note that $\tau_{SF}$ would
be equal to the gas depletion timescale if there was no accretion
of fresh gas from cooling, no ejection of gas by SN feedback and
no recycling of gas to the ISM by dying stars.) Since the treatment of
gas cooling in halos is unchanged, accretion rates of cold gas onto
disks are also essentially unchanged, so an increase in the SF
timescale $\tau_{SF}$ leads to larger cold gas masses in disks and
vice versa. The new SF laws depend on galaxy properties such as the
radius, stellar mass and gas metallicity in different ways from the
old laws. Furthermore, all of the new SF laws are non-linear in 
cold gas mass, at least in the low gas density regime, so that
$\tau_{SF}$ always increases if $M_{\rm cold}$ becomes low
enough, in contrast to the original SF laws, for which $\tau_{SF}$ is
independent of the remaining gas mass. As a result, the changes in gas
content due to changing the SF law themselves depend on the galaxy
mass and redshift {(see Appendix~\ref{App:SFR} for the change in SFR
with galaxy mass and redshift for different SF laws).} 

Starbursts triggered by disk instabilities have a 
bearing on gas content in the Bow06 model. 
Changing the SF law can lead to changes in the total disk mass, 
making the disk more prone to instabilities if 
the total mass is larger. 
Following an instability, all of the
gas is consumed in a starburst. Thus, paradoxically, longer SF
timescales in disks can lead to {\em lower} final gas contents in some
cases. Starbursts in the Bau05 model are triggered only
by galaxy mergers, and the frequency of these is essentially
unaffected by changes in the gas content of galaxies.

In the Bau05 model, the SF timescales are in most cases shorter with
the new SF laws, and so the CMF is also lower at most gas masses. For
the Bow06 model, the results of changing the SF law are more mixed,
although the CMF is generally lower at intermediate gas masses. 
An interesting effect appears at low gas masses,
where in all cases, the number density is larger
than with the original SF laws, bringing the
models closer to the observational data at $z=0$. This results from
the new SF laws being non-linear in the cold gas mass, so that the
gas is depleted less rapidly when the mass becomes very small 
than with the original (linear) SF laws. 
The effects of halo mass resolution are more severe for the
Bow06 model, which uses N-body merger trees
from the Millennium simulation. This is what causes the turndown
in the CMF for $M_{\rm cold} \lsim 10^{9}
h^{-2}\, M_{\odot}$.

The high-mass end of the CMF in the Bow06 model 
evolves significantly from $z=2$ to
$z=0$ due to the longer SF timescales at lower 
redshifts (see Eq.~$2$). With the new SF laws, this 
evolution is weaker, due in part to the 
higher frequency of disk instabilities triggering starbursts
which consume the gas.

The evolution of the CMF is therefore a useful diagnostic to
distinguish between the SF laws, since they lead to changes 
that are differential with galaxy mass.  

\subsection{Global cold gas density evolution}

\begin{figure}
\begin{center}
\includegraphics[trim = 1mm 1mm 1.5mm 1.5mm,clip,width=0.45\textwidth]{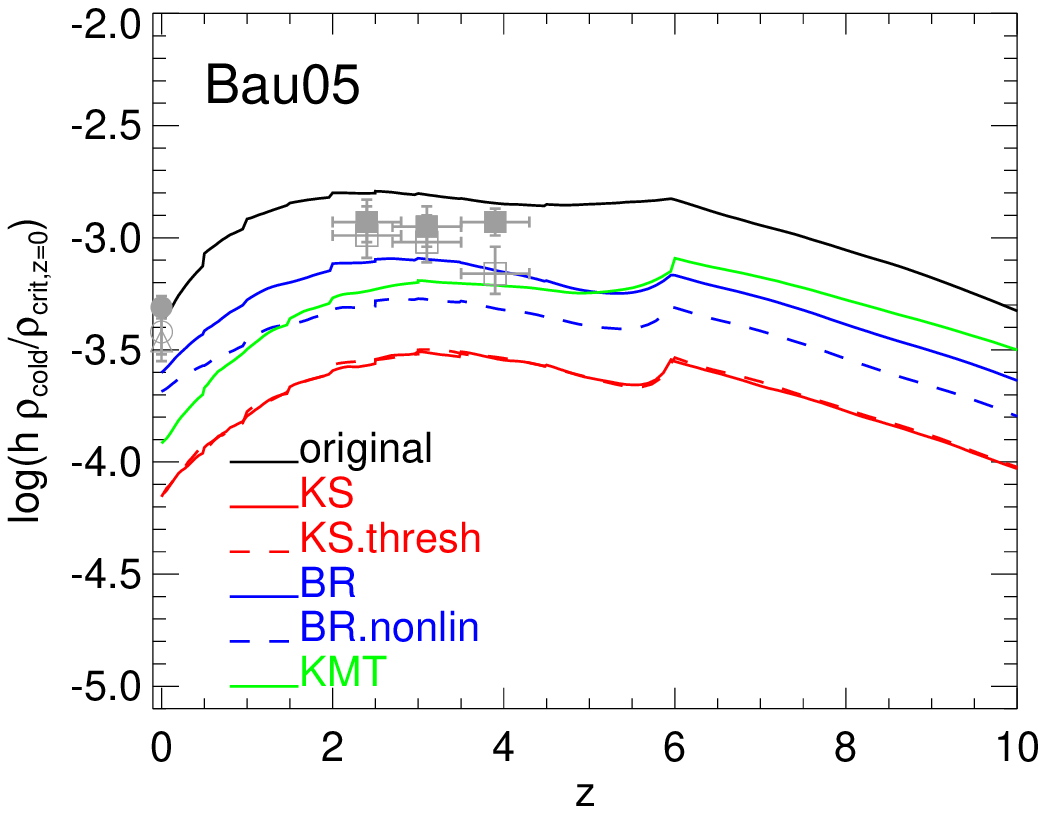}
\includegraphics[trim = 1mm 1mm 1.5mm 1.5mm,clip,width=0.45\textwidth]{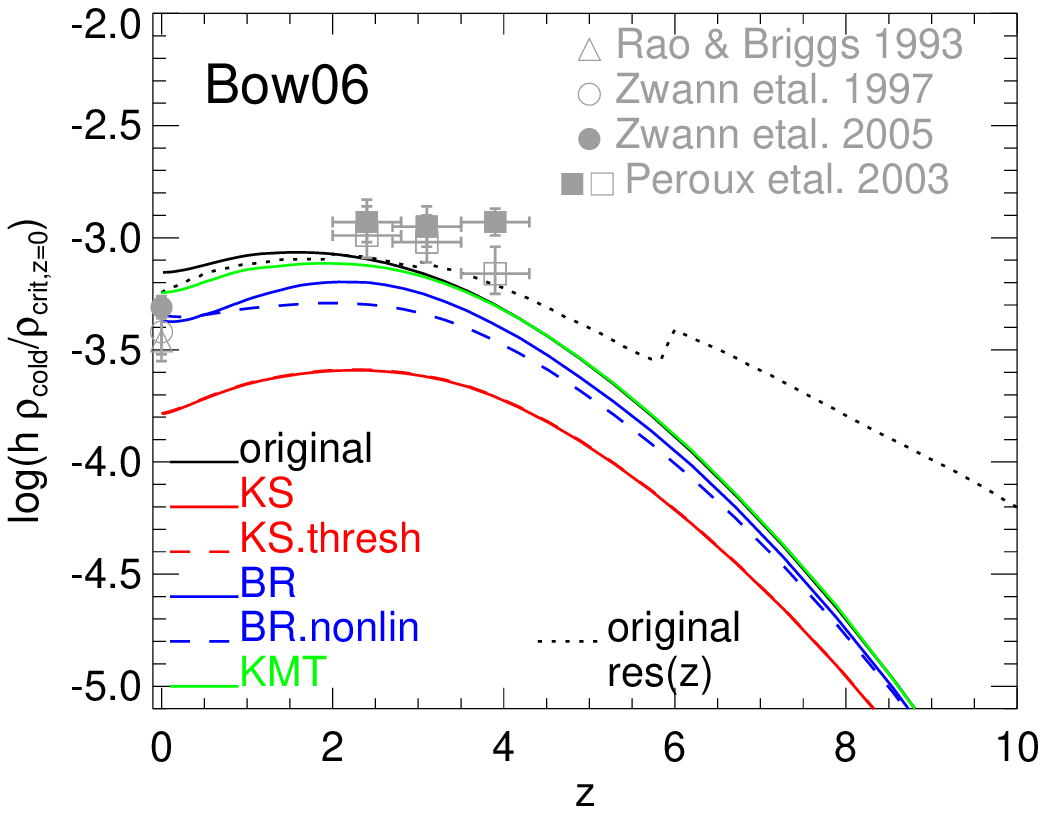}
\caption{The cold gas mass density in the universe, $\rho_{\rm cold}$,
in units of the critical density at $z=0$, $\rho_{\rm crit,z=0}$, as a
function of redshift.  The results for the original Bau05 and Bow06
models and with the new SF laws are shown in the top and bottom panels
respectively.  Grey symbols show observational results from
\citet[open and filled circles, respectively]{Zwaan97,Zwaan05},
\citet[open triangles]{Rao93} and \citet[open and filled
squares]{Peroux03}. For the latter, open symbols indicate the cold gas
density inferred from damped Ly$\alpha$ systems, while filled symbols
include a correction for gas clouds with lower column density than
originally detected. The dotted line in the bottom panel shows the
effect of changing the halo mass resolution from the default values,
as discussed in the text.}
\label{CGMevo}
\end{center}
\end{figure}

The evolution of the global cold gas density, $\rho_{\rm cold}$, is
shaped by the same processes as the CMF.  Fig.~\ref{CGMevo} shows
$\rho_{\rm cold}$ as a function of redshift 
for the Bau05 (top panel) and Bow06 (bottom panel) models and the
variants with new SF laws.  For reference, grey symbols show
different observations as listed in the legends.  
In general, all the new SF laws 
predict lower $\rho_{\rm cold}$ than the original
models across the whole redshift range.

In the Bau05 model, the reduced $\rho_{\rm cold}$ with the new SF laws 
is due to the shorter quiescent SF timescales in low and
intermediate cold gas mass objects (see right-hand panels of
Fig.~\ref{taus}). In general, the shorter 
the quiescent SF timescale (on average) the 
smaller $\rho_{\rm cold}$ compared to the 
original model.
Note the jump of $\rho_{\rm cold}$ at $z\approx
6$ in the Bau05 model is due to the assumed reionization redshift
$z_{\rm reion}=6$ ($\S 6.1$).

On the other hand, in the Bow06 model the BR, BR.nonlin and the KMT SF
laws at very high redshift (i.e. $z\ge6$) give similar $\rho_{\rm
cold}$ to the original model, mainly due to the compensation between
quiescent and burst SF activity. The BR, BR.nonlin and the KMT SF laws give 
reasonable agreement with the observed $\rho_{\rm cold}$ at $z=0$, 
improving over the original model.

The Bow06 model shows a more rapid decrease of $\rho_{\rm cold}$ with
increasing redshift than the Bau05 model. This results in part
from the stronger SNe feedback in the Bow06 model, but is also due 
to the difference in the mass resolution of the halo merger
trees used in the two models (see \S\ref{sec:galform}). 
The dotted line in the bottom 
panel of Fig.~\ref{CGMevo} shows the effect on the standard Bow06
model of using Monte Carlo trees with a minimum mass $10^{10} h^{-1}
M_{\odot}(1+z)^{-3}$ instead of N-body trees with a 
fixed mass resolution (shown by the solid
black line). The effect of improving the halo mass resolution is 
larger at high redshifts. 

\subsection{Gas-to-luminosity ratios of galaxies}
\label{sec:gasfrac}

\begin{figure}
\begin{center}
\includegraphics[trim = 2mm 1mm 1.5mm 0mm,clip,width=0.45\textwidth]{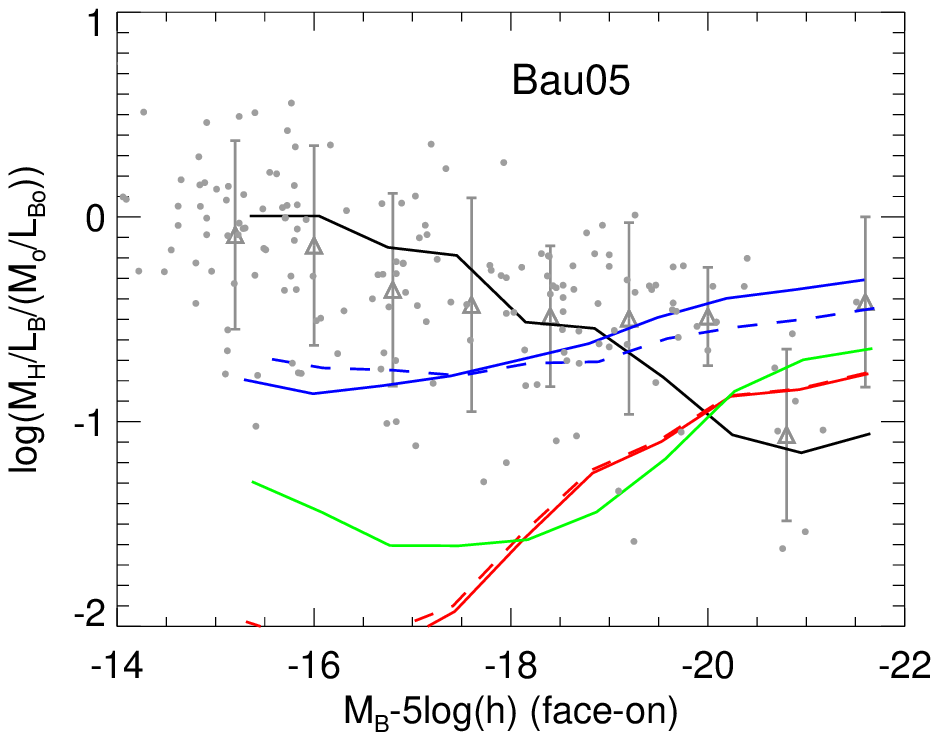}
\includegraphics[trim = 2mm 1mm 1.5mm 0mm,clip,width=0.45\textwidth]{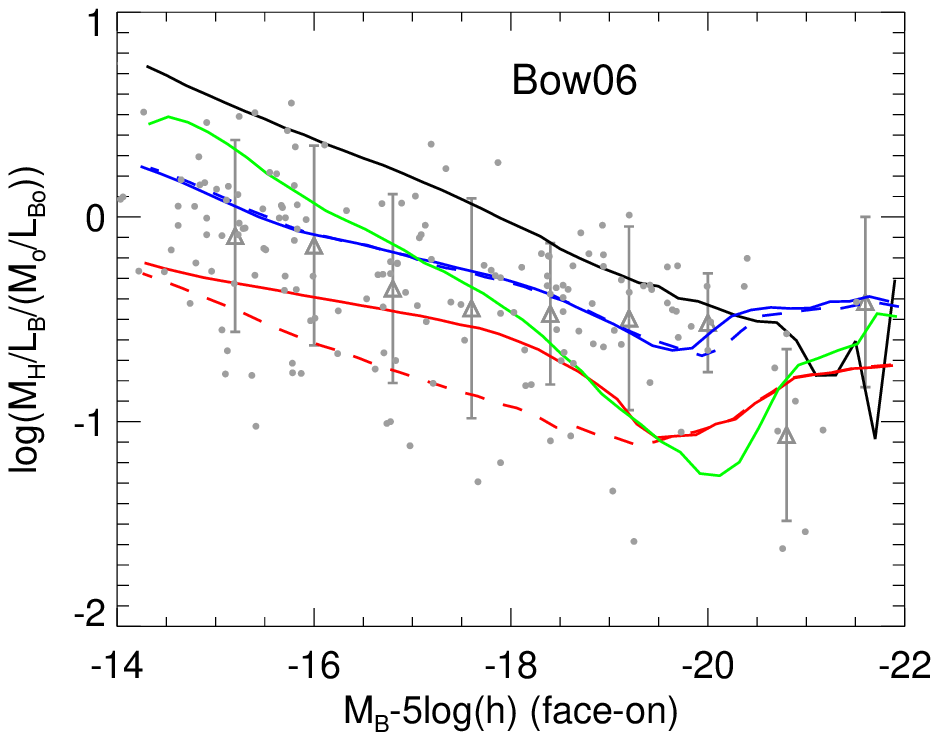}
\caption{Hydrogen gas mass to luminosity ratios, $M_{\rm H}/L_{\rm B}$, as a
function of $B$-band magnitude for late-type galaxies. Late-type 
galaxies in the model are those with a bulge-to-total luminosity 
in the $B$-band of $B/T\le 0.4$. The Bau05 and
Bow06 models are shown in the top and bottom panels respectively.
The line colours and styles which represent the different SF laws are
the same as in Fig.~\ref{LF}.  For clarity we only show the median of
each model. Observational results from \citet{Huchtmeier88} and
\citet{Sage93} are shown as small gray symbols.  The triangles with
errorbars show the median and the $10$ and $90$ percentiles for the
observations. In order to compare directly with the observations, we
assume that $74$\% of the cold gas mass predicted by the models is
hydrogen.}
\label{CMFracs}
\end{center}
\end{figure}

Fig.~\ref{CMFracs}
shows the gas-to-luminosity ratio as a function of absolute magnitude
in the $B$-band at $z=0$ for late-type galaxies (those with bulge-to-total 
luminosity in the $B$-band, $B/T\le
0.4$ in order to make 
a fair comparison with observations), for the original models and with the new SF laws. 

The Bau05 model predicts gas-to-luminosity ratios in good
agreement with the observations, as these data were used as
the main constraint on the parameters in the original SF recipe
(eqn.\ref{SFRold}). The new SF laws in the Bau05
model predict gas-to-luminosity ratios which increase with luminosity,
contrary to the observed tendency. This is due to the new SF laws
predicting generally shorter quiescent SF timescales in low and
intermediate mass galaxies but longer timescales in high mass
galaxies, when compared to the original model (see Appendix~\ref{App:SFR}).

The original Bow06 model predicts gas-to-luminosity ratios which are
in poor agreement with the observations, except at the highest
luminosities. However, the new SF laws
generally predict lower gas-to-luminosity ratios for galaxies fainter
than $M_B-5 \log( h) \approx -20$, due to shorter quiescent SF
timescales, bringing them into better agreement with the observations.

\section{The evolution of galaxies in the SFR vs. stellar mass plane}
\label{SFHgals}

{The level of SF activity in galaxies at any
redshift is dependent on the available cold gas and 
the timescale on which the gas is consumed 
(which is set by the choice of SF law). Here we 
examine the influence of different SF laws on the distribution of galaxies 
in the SFR-stellar mass (SFR-$M_{\star}$) plane and its evolution 
with redshift. The distribution of galaxies in this plane 
offers a way to constrain the form of the SF law.
}

\citet{Brinchmann04} found that local 
star-forming galaxies lie on a relatively narrow region in the
SFR-$M_{\star}$ plane. \citet{Noeske07}, \citet{Elbaz07} and
\citet{Daddi07} extended this work to higher
redshifts, finding that the ``main sequence'' of star-forming galaxies
in the SFR-$M_{\star}$ plane remains in place up to at least $z\sim 2$
with a similar slope to that at $z \sim 0$. The zero-point evolves 
strongly with redshift. \citet{Santini09} found
evidence for a bimodal distribution of galaxies in the SFR-$M_{\star}$
plane, with a ``passive'' sequence below that of actively
star-forming galaxies. Below, we investigate whether or not the models are
able to explain the observations, and to what extent the
distribution of galaxies in the SFR-$M_{\star}$ constrains 
the SF law.

\subsection{The local star formation rate-stellar mass plane }

\begin{figure*}
\begin{center}
\includegraphics[width=0.45\textwidth]{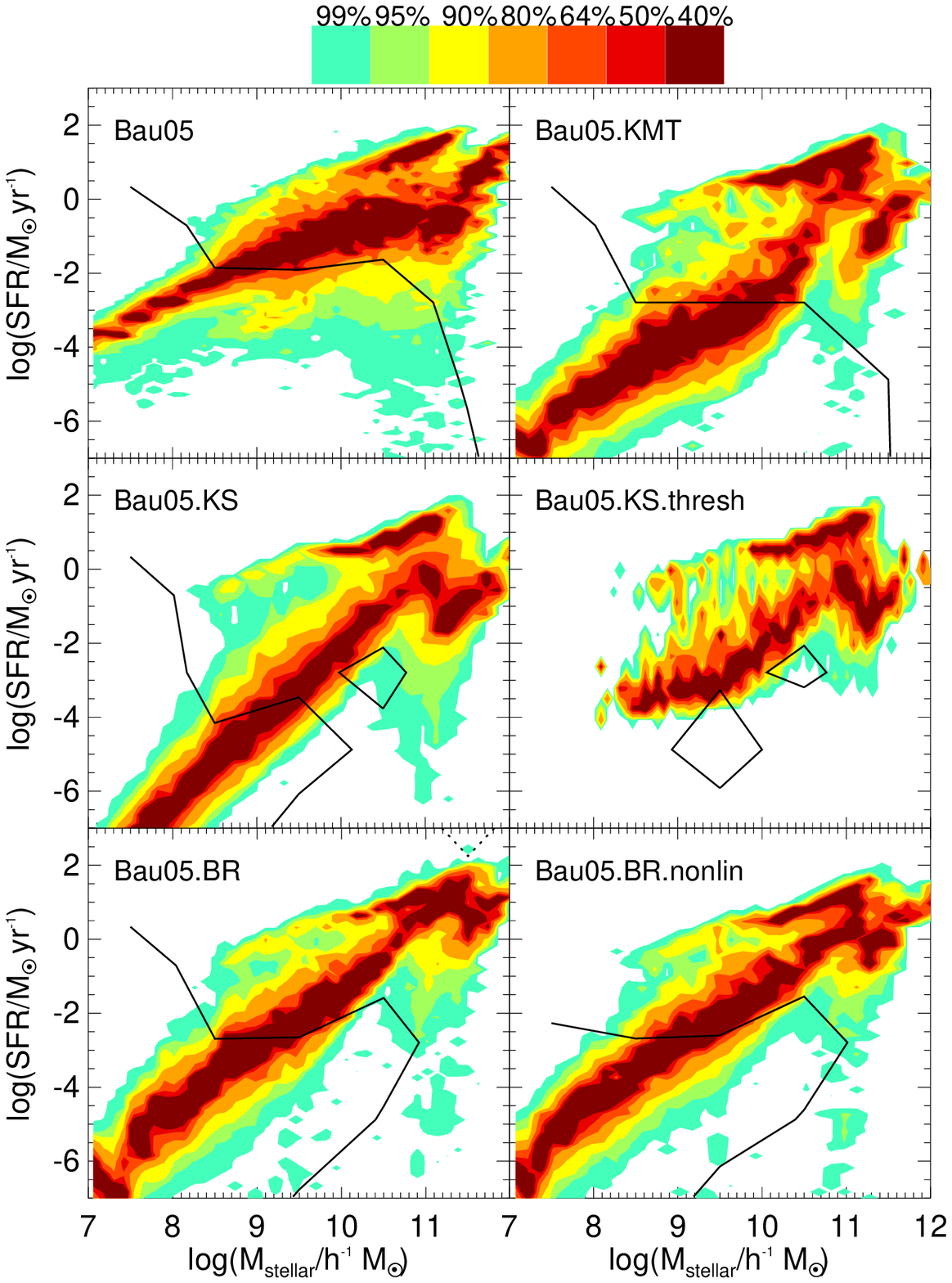}
\includegraphics[width=0.45\textwidth]{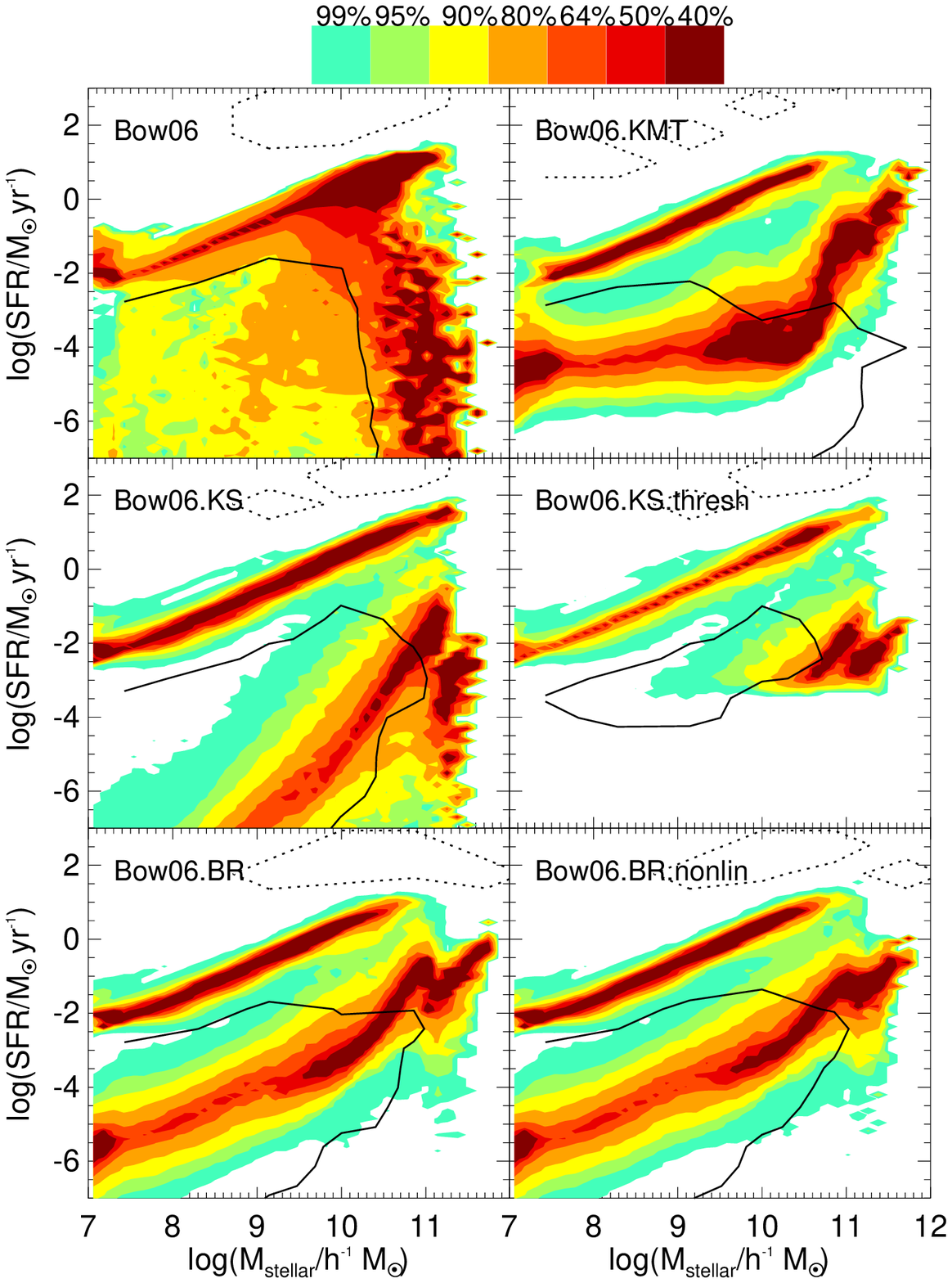}
\caption{
Distribution of galaxies in the SFR vs. stellar
mass plane at $z=0$ for the Bau05 and the Bow06 models and
their variants with the new SF laws (left and right-hand sets of
panels respectively). Each panel corresponds to a different model, as
labelled. The coloured contours show the regions within which
different fractions of the galaxies lie for given stellar mass,
volume-weighted and normalized in bins of stellar mass, with the scale
shown by the key. We show, using dotted and solid black 
contours, the regions in which starburst and
satellite galaxies respectively make up more than $50$\% of the galaxy
population.}
\label{SFRMsevo1}
\end{center}
\end{figure*}

Fig.~\ref{SFRMsevo1} shows the predicted distribution of galaxies in
the SFR-$M_{\star}$ plane at $z=0$ for the Bau05 and Bow06
models and for variants using the new SF laws.  The
colour shading indicates the regions within which $40$\% to $99$\% of
the model galaxies fall in the SFR-$M_{\star}$ plane, when the
distributions are volume-weighted and normalized in bins of stellar
mass.  We also show, using solid and dotted black contours, the
regions of the plane where satellite and burst galaxies respectively
constitute more than $50$\% of the population within cells in the
SFR-$M_{\star}$ plane. Here we define burst galaxies as those in
which the SB mode accounts for at least $20$\% of their total SFR.
Most of the galaxies ($>99\%$ when integrated over the whole
SFR-$M_{\star}$ plane) are undergoing quiescent SF. Some of these
contours break up into ``islands'', but this is an
artifact of some cells in the SFR-$M_{\star}$ plane 
containing a small number of galaxies.

We see from Fig.~\ref{SFRMsevo1} that all of the models apart from the
Bow06 model show two sequences in the SFR-$M_{\star}$ plane,
although the shapes and widths of these sequences vary significantly
depending on the SF law. We will call the upper sequence with higher
SFRs at a given $M_{\star}$ the ``active'' or ``star-forming''
sequence, and the lower sequence with lower SFRs the ``passive''
sequence. In the original Bow06 model there appears to be only an
active sequence, but there is a large spread of galaxies down to lower
SFRs. The two sequences appear more distinct for the variants of the
Bow06 model with the new SF laws than for the Bau05 model and its
variants. For the Bow06 model and variants, the active SF sequence
extends down to the lowest stellar masses plotted, while for the Bau05
model and its variants, the active sequence is only present for higher
stellar masses. We now examine these differences in more detail.

In the Bau05 model and its variants, there are comparable
numbers of active and passive galaxies for $M_{\star}
\gsim 10^{10} h^{-1}\Msol$. At lower masses most of the galaxies
are on the passive sequence, with the active sequence completely
disappearing for $M_{\star} \lsim 10^{9} h^{-1}\Msol$. The passive
sequence is generally broader than the active one. The slope of
the passive sequence is generally steeper with the new SF laws,
producing lower SFRs at $M_{\star} \lsim 10^{10}\, h^{-1}\,
M_{\odot}$.  This is closely related to the lower gas content in
these galaxies, as seen in Fig.~\ref{CMFracs}. The exception is the
KS.thresh law, where the passive sequence disappears at low SFR and
low $M_{\star}$. This results from the cut-off in SF for galaxies with
gas surface densities below the critical density (see
Eq.~\ref{gascut}).  Indeed, $70$\% of all galaxies in this model with
$M_{\star}>10^7\,h^{-1}\rm M_{\odot}$ have $\rm SFR=0$. However,
in galaxies with $M_{\star}>10^9\, h^{-1}\rm M_{\odot}$, this
percentage falls to $30$\%.  This means that the critical density 
cut-off completely switches off SF activity in most low mass 
galaxies at $z=0$.

\begin{figure}
\begin{center}
\includegraphics[trim = 3mm 1.5mm 1.5mm 1.5mm,clip,width=0.45\textwidth]{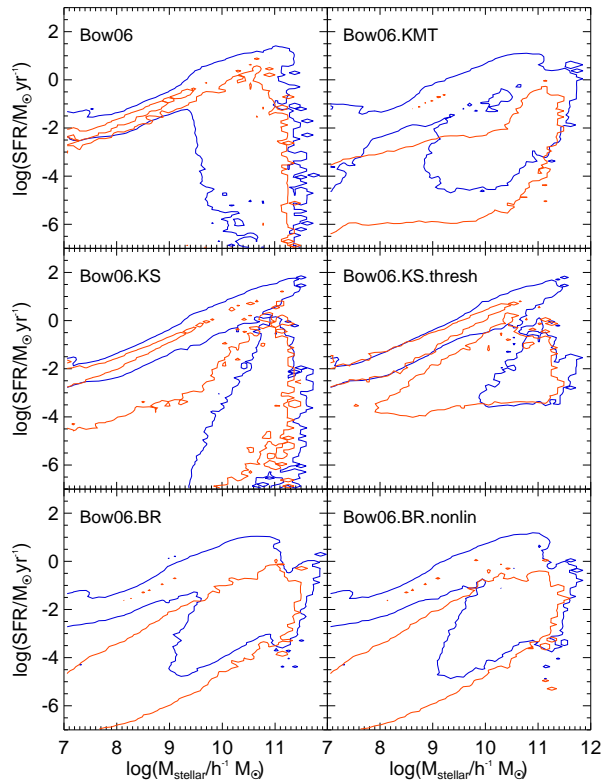}
\caption{SFR vs. stellar mass plane for galaxies at $z=0$ in the Bow06
model with the different SF laws tested. The contours outline the
$95$\% percentile regions for satellite (red) and central (blue)
galaxies, after normalizing in bins of stellar mass.}
\label{SFRMs_sc}
\end{center}
\end{figure}

For the variants of the Bow06 model the passive
sequence appears generally narrower and more offset from the active
one, compared to the Bau05 model variants; furthermore, there seems 
to be more of dependence of the shape on the SF law. 
As for the Bau05.KS.thresh model, the passive sequence disappears 
completely at low SFRs in the Bow06.KS.thresh model.

In order to gain further insight into the nature of the upper and lower
sequences in the SFR-$M_{\star}$ plane, we split the sample into
central and satellite galaxies. Fig.~\ref{SFRMs_sc} shows the $95$\%
percentile contours after normalization in bins of stellar mass, for
central (blue contours) and satellite galaxies (red contours) for the
Bow06 model and its variants. Satellite galaxies are found
predominantly on the lower sequence, while the upper sequence is
dominated by central galaxies, as expected from the solid black
contours in Fig.~\ref{SFRMsevo1}. This implies that the inflow of
newly cooled gas onto central galaxies is a key process in shaping the
``active''sequence, while the rate at which
the cold gas reservoirs in satellite galaxies are consumed shapes the
lower, ``passive'' sequence. In satellite galaxies, 
the new SF laws applied to the Bow06 model typically produce 
longer SF timescales (see Appendix~\ref{App:SFR}). This is a 
consequence of the new SF laws being non-linear in cold gas mass, 
hence the SF timescale becomes long when the gas mass becomes small, 
which happens when galaxies no longer accrete newly cooled gas. 
Therefore satellite galaxies consume their
gas reservoirs more slowly than in the original model, which results
in larger cold gas masses and higher SFRs at later epochs, and allows
the galaxies to remain on a well defined sequence for longer. 
In contrast, with the old SF recipe the gas reservoirs of satellites decline
exponentially with time.

The change in the slope of the passive sequence at $M_{\star}\sim
10^{10} h^{-1} M_{\odot}$ in the Bow06.BR, Bow06.BR.nonlin and
Bow06.KMT models is due to the change in the dominant regime in the
$\Sigma_{\rm SFR}-\Sigma_{\rm gas}$ relation. 
Low mass galaxies are predominantly in the
low $\Sigma_{\rm gas}$ regime, for which most of the hydrogen is in
the atomic phase, and the SFR has a steep dependence on $\Sigma_{\rm
gas}$. Indeed, the steeper the relation in the $\Sigma_{\rm
SFR}-\Sigma_{\rm gas}$ plane at low $\Sigma_{\rm gas}$, the flatter 
the passive sequence at low stellar masses (see Fig.~\ref{SFRMsevo1}
for the KMT SF law). However, this simple picture only holds for
satellite galaxies, since centrals have a constant supply of newly 
cooled gas which shapes the upper sequence in the SFR-$M_{\star}$ plane, 
making it insensitive to the exact choice of SF law.

In the Bau05 model the satellite and central galaxy sequences overlap
more due to the longer timescale for gas to be reincorporated
into the host halo after ejection by SNe, compared with
the Bow06 model (see $\S 2$). This is particularly important at low
stellar masses, for which SNe feedback is very efficient at ejecting
gas which has previously cooled. The long reincorporation timescale
means that star formation is nearly shut down in most low-mass central
galaxies. Hence the active SF sequence disappears at low
$M_{\star}$. The passive sequence at low $M_{\star}$ therefore
includes both satellite and central galaxies. We confirmed this 
by carrying out the exercise of applying the Bow06
prescription for the gas reincorporation timescale in the Bau05 model,
and found an active sequence which looks similar to that in the
Bow06 model. 
The slope and dispersion of the active SF 
sequence are insensitive to the details of the SNe
feedback (or the SF law; see Fig.~\ref{SFRMsevo1}), since making feedback 
stronger or weaker does not change its shape. We conclude that the
reincorporation timescale of the ejected gas is the main process
controlling the form of the active SF sequence, {while the passive 
sequence is mostly dependent on the choice of SF law. This relation 
offers a new way to constrain the SF law.}

\subsection{Comparison with observations of the SFR-$\rm M_{\star}$
  plane at $z=0$}

\begin{figure}
\begin{center}
\includegraphics[width=0.35\textwidth]{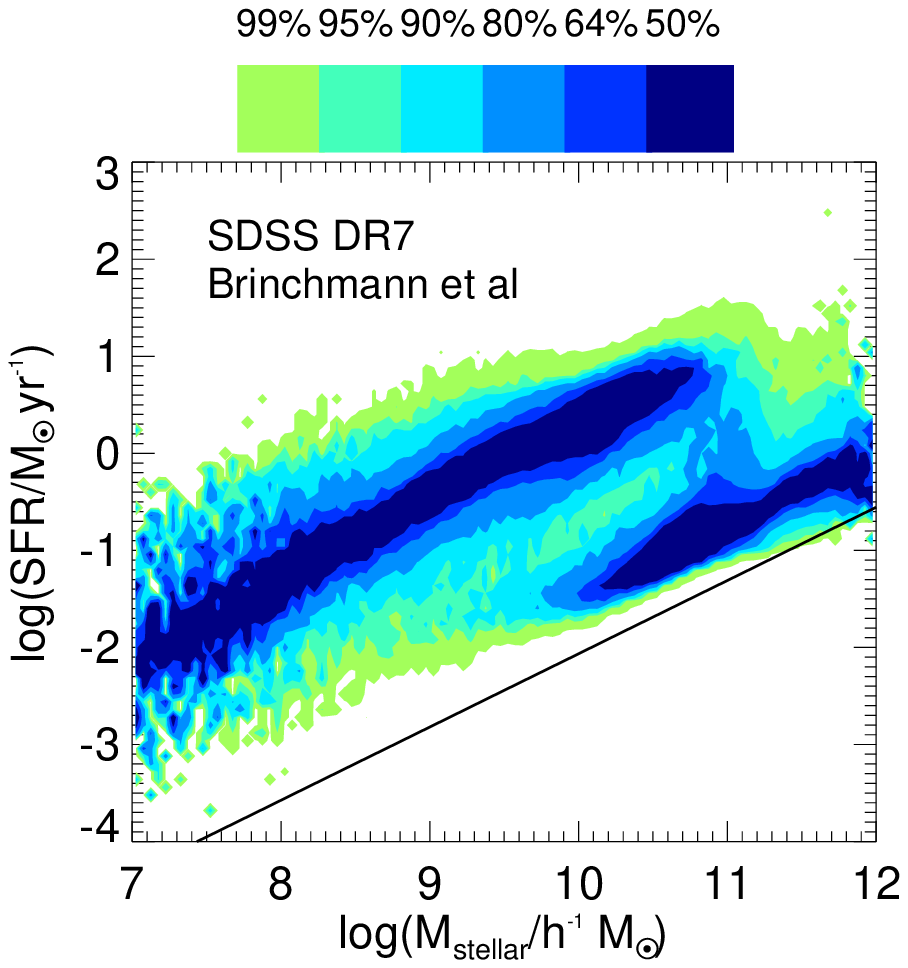}
\includegraphics[width=0.49\textwidth]{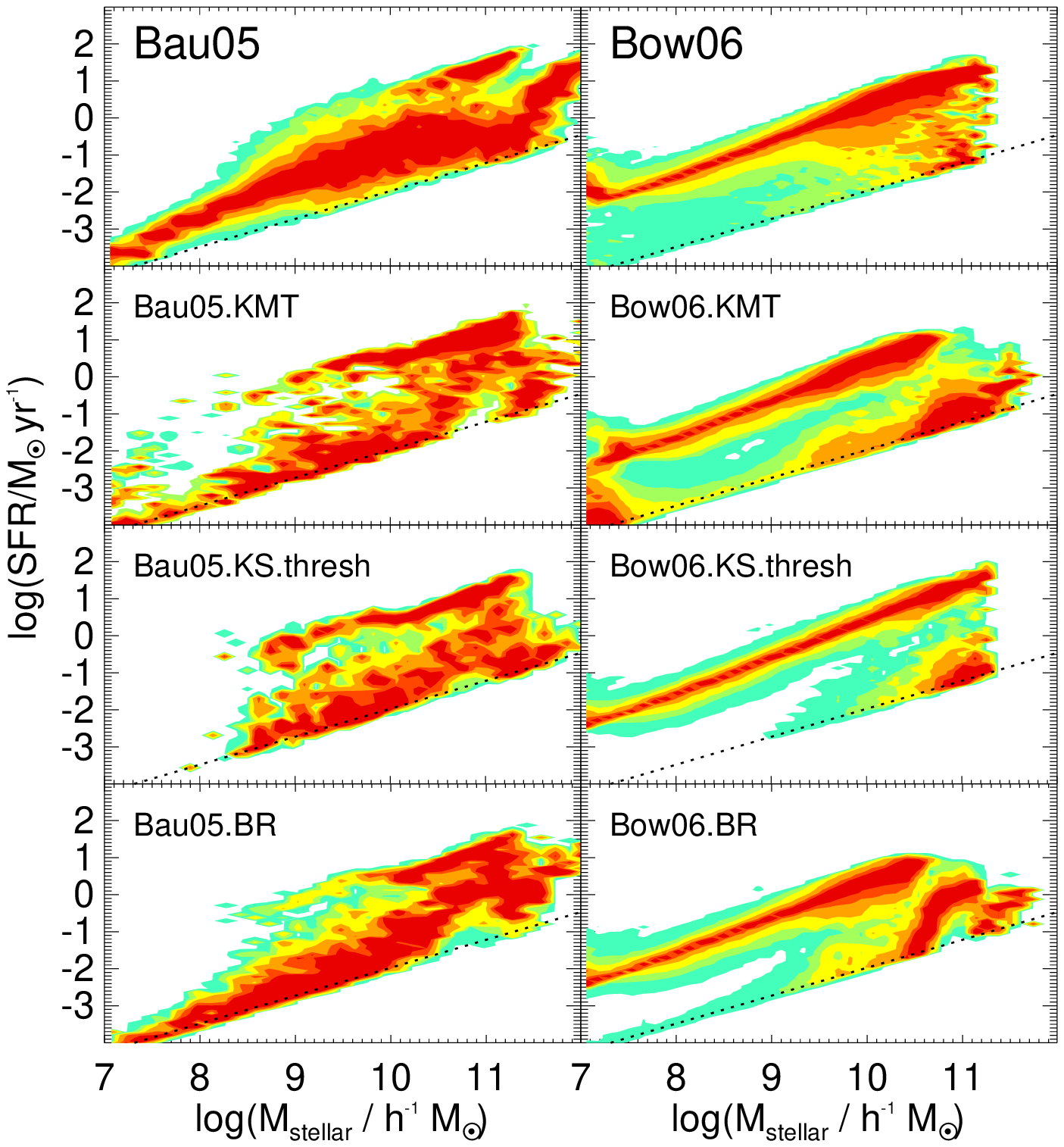}
\caption{{\it Top panel:} Observed distribution of galaxies in the SFR vs. stellar mass
plane for the SDSS DR7 spectroscopic sample of galaxies, updating the
analysis by \citet{Brinchmann04}.  The distributions are volume-weighted
and normalized in bins of stellar mass as in Fig.~\ref{SFRMsevo1},
and the shading shows the regions within which different
fractions of galaxies lie at each stellar mass. The black line shows
the effective SFR sensitivity, which depends on stellar
mass (Brinchmann, priv. com.). {\it Bottom panel} 
The same as the top panel
 for model galaxies at $z=0.1$ after applying the cuts
used by \citet{Brinchmann04} in the SDSS DR7 sample, for different
models as labelled in each panel. For reference, the dotted lines show
the sensitivity limit of the SFR estimates in the SDSS DR7. 
The shading is as in the top panel. 
{The KS and BR.nonlin 
models are not shown due to their similarity with the KS.thresh 
and BR models respectively.}}
\label{SFRMsmodel-SDSS}
\end{center}
\end{figure}

Indications of multiple sequences in the SFR-$M_{\star}$
plane have been found in the local Universe by \citet{Brinchmann04},
and at high redshifts by \citet{Damen09}, \citet{Santini09} and
\citet{Rodighiero10}.  However, all of these studies are affected to
some extent by selection effects against the inclusion of galaxies 
with low SFRs and/or low stellar masses, and these must be accounted 
for in any comparison with models. We start by comparing with 
observations of local galaxies in this subsection, and then consider higher
redshifts in the next subsection.

In the top panel of Fig.~\ref{SFRMsmodel-SDSS} we show the observed distribution of galaxies
in the SFR-$M_{\star}$ plane based on the SDSS Data Release 7
(DR7)\footnote{Data was downloaded from the public webpage
\texttt{http://www.mpa-garching.mpg.de/SDSS/DR7/}.}, which corresponds
to an update of the \citet{Brinchmann04} analysis. Stellar masses are
determined from spectra and broad band magnitudes 
following \citet{Kauffmann03}, while SFRs are
derived primarily from the $H_{\alpha}$ emission line following
\citet{Brinchmann04}. We refer the reader to these papers for further
details. We have constructed contours of the
distribution for the subsample of galaxies in the spectroscopic
catalogue which have estimated SFRs, after volume-weighting and
normalizing in bins of stellar mass. As in Fig.~\ref{SFRMsevo1}, the
colour shading indicates the regions within which different fractions
of galaxies lie for a given stellar mass. The solid black line shows
the approximate SFR sensitivity limit in the SDSS analysis (Brinchmann,
priv. comm.). The minimum measurable SFR depends on stellar mass both
because it depends on detecting spectral features above the stellar
continuum, and because galaxies of higher $M_{\star}$ in the SDSS
sample tend to lie at larger distances. Since the SDSS estimates are
based on a \citet{Chabrier03} IMF for the stellar masses, and a
\citet{Kroupa01} IMF for the SFRs, we rescale the stellar masses in
the SDSS sample by a factor of $0.89$ and the SFRs by a factor of
$1.1$ (\citealt{Bell03}) to correspond to the use of 
a \citet{Kennicutt83} IMF for quiescent star formation in the 
models. (In the Bau05 model starbursts form stars with a top-heavy IMF, 
but this only affects a small fraction of the galaxies in the SFR-$M_{\star}$
plots. SB galaxies are typically located above the colour contours at
higher SFRs, see dotted lines in Fig.~\ref{SFRMsevo1} and
Fig.~\ref{SFRMsevoBow06}).

To make the comparison with the SDSS data simpler and fairer,
we plot the model predictions in the SFR-$M_{\star}$
plane again in the bottom panel of Fig~\ref{SFRMsmodel-SDSS}, but this time imposing some
additional selections to better match the SDSS sample. Since the
median redshift of the SDSS spectroscopic sample is around $z=0.1$, we
create a sample of model galaxies for $z=0.1$, and then apply the
apparent magnitude limit $r<17.77$ of the SDSS spectroscopic
sample. We also apply the SFR sensitivity limit of the SDSS sample
shown in the top panel of Fig.~\ref{SFRMsmodel-SDSS}.  Fig~\ref{SFRMsmodel-SDSS} shows the
SFR-$M_{\star}$ plane for the Bau05 (left panels) and the Bow06 (right
panels) models at $z=0.1$ after applying these cuts. {Note that we 
do not show the BR.nonlin and KS models given their similarity to 
the BR and KS.thresh, respectively.}

For the SDSS sample plotted in Fig.~\ref{SFRMsmodel-SDSS}, we see that most
galaxies in the mass range $10^7 < M_{\star} \lsim 10^{10}
h^{-1}\Msol$ lie on a single star-forming sequence. At high masses
$M_{\star} \gsim 10^{10} h^{-1}\Msol$, a second sequence appears,
roughly parallel to the first but offset below it. Comparing with
the model distributions in 
Fig.~\ref{SFRMsmodel-SDSS}, the general distribution of galaxies in the
SFR-$M_{\star}$ plane seems better matched by the variants of the
Bow06 model rather than by any of the variants of the Bau05 model. 
All of the variants of the Bow06 model
produce an upper star-forming sequence covering the whole mass range
$10^7 < M_{\star} \lesssim 10^{11} h^{-1}\Msol$, with similar slope to the
observed one, though with a smaller dispersion, particularly at low
masses. The original Bow06 model does not reproduce the second
sequence seen in the SDSS at high masses, but most of the variants
with new SF laws do produce a similar feature. On the other hand, in 
the Bau05 model, the distribution of galaxies in the SFR-$M_{\star}$ 
plane appears rather different from the
observed one. All the variants show an upper sequence of star-forming
galaxies, which has a similar slope and amplitude to that seen in
SDSS. However, this does not extend to low masses, unlike in the SDSS. 
The most prominent feature for the
Bau05 model and its variants is the broad ``passive'' sequence lying
below the ``active'' star-forming sequence, and extending down to the
lowest masses plotted. This does not correspond to what is seen in
the SDSS sample, where all low-mass galaxies appear to lie close to
the ``active'' star-forming sequence. The main origin of these
differences between the Bau05 and Bow06 model for low stellar masses
is the difference in the timescale for the reincorporation into halos
of gas which has been ejected by supernovae (see $6.1$). 

The passive sequence seen in Fig.~\ref{SFRMsevo1} for the variants of
the Bow06 model falls below the SFR sensitivity
limit of the SDSS study at lower masses, $M_{\star} \lsim 10^{10}
h^{-1}\Msol$.  However, observational indications of a passive galaxy
sequence at low masses have been reported by \citet{Woo08} for local
group dwarf galaxies. These objects follow a distinct SF sequence with
much lower SFRs than the main star-forming sequence seen in the SDSS
sample, similar to our predictions for passive galaxies in the Bow06
model variants with the new SF laws.  Similarly, \citet{Skillman03}
reported SFRs and stellar masses for galaxies in the Coma cluster
which agree with the lower sequence seen in the top panel of Fig.~\ref{SFRMsmodel-SDSS}, and
also overlap with our prediction for massive satellite galaxies in the
Bow06.BR, Bow06.BR.nonlin and Bow06.KMT models (Fig.~\ref{SFRMs_sc}).

\subsection{The SFR-$\rm M_{\star}$ relation at high redshift}

\begin{figure}
\begin{center}
\includegraphics[trim = 3.5mm 3.5mm 1.5mm 1.5mm,clip,width=0.49\textwidth]{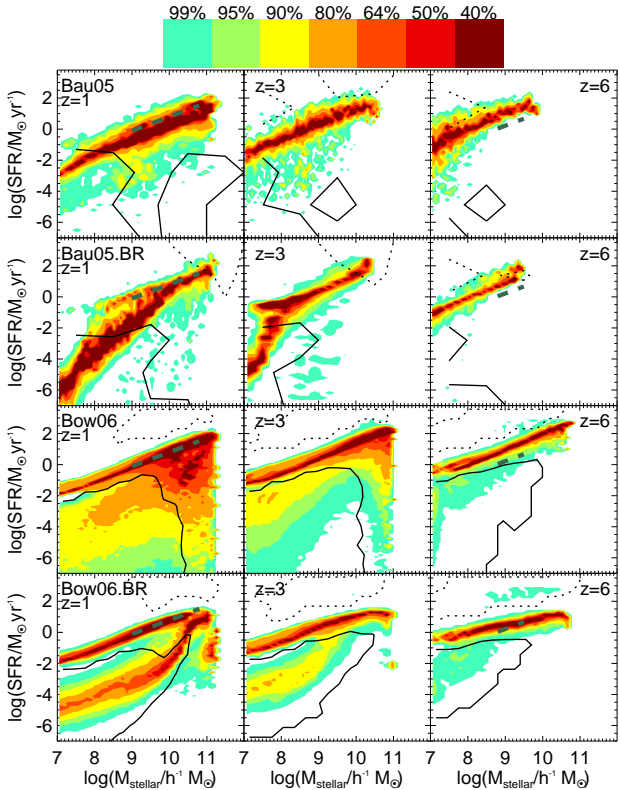}
\caption{Distribution of galaxies in the SFR vs stellar mass 
plane at $z=1$ (left column), $z=3$ (middle column) and $z=6$ (right column) 
{for the Bau05 (top panel), the Bau05.BR (second panel), the 
Bow06 (third panel) and 
the Bow06.BR models (bottom panel).} The shading shows the
distribution of galaxies in this plane, normalized in bins of stellar
mass.  The thick dashed straight lines show the observations of the ``SF
sequences'' reported by \citet{Elbaz07} at $z\sim 1$ and
\citet{Stark09} at $z\sim 6$, and are plotted over the mass ranges 
probed by the observations. We also show regions in 
which starburst and satellite galaxies make up more than
$50$\% of the population using dotted and solid black contours,
respectively.}
\label{SFRMsevoBow06}
\end{center}
\end{figure}

Evidence for a star-forming sequence in the SFR-$M_{\star}$ plane 
has also been found at high redshifts. The
slope of the observed relation seems to be constant over the redshift
range probed, while the zero point evolves strongly from $z\approx0$
to $z\sim 4$ \citep{Elbaz07,Daddi07,Perez-Gonzalez08}, but appears
roughly constant for $4 \lsim z \lsim 6$ \citep{Stark09}. However, we
caution that the selection effects in some of the high-redshift
samples are very strong, since some are selected on the basis of
SFRs. Furthermore, the observational estimates of SFR and stellar mass
typically become more uncertain at higher-z, due to the more limited
data available and the effects of dust \citep{Stringer10}. 

{Both models in all of their variants 
predict an active SF sequence in good agreement with observations. 
The features of the models in the SFR-$M_{\star}$ 
plane at $z=0$ (see Fig.~\ref{SFRMsevo1}) are preserved 
up to $z\approx 2$. At $z>2$, the passive sequence 
 starts to disappear, implying that at higher redshifts most of 
the galaxy population is undergoing vigorous SF 
and lies on the active SF sequence. 
To illustrate this general 
behaviour, Fig.~\ref{SFRMsevoBow06} shows the predicted 
distribution of galaxies in the SFR-$M_{\star}$ 
at
three different redshifts for the Bau05 and the Bow06 models, together
with the BR variant. Also shown are the medians of the observed
relations at different redshifts, plotted as thick dashed lines}\footnote{{ 
Since in the observational samples a universal
Salpeter IMF was assumed when estimating SFRs and stellar masses, 
we scale their stellar masses and SFRs by factors
of $0.5$ and $0.91$ respectively (e.g. \citealt{Bell03}), to
correspond to a \citet{Kennicutt83} IMF.}}. 

\citet{Santini09} found that bimodality in the
SFR-$M_{\star}$ plane is clearly present in massive galaxies at
$z\lsim 1$, but becomes weaker as redshift increases, being almost { absent} 
at $z\sim 2$. 
This appears consistent with the predictions of the Bow06 model with the new SF laws, 
but is probably less so with predictions from the Bau05 model. However,
the SFR sensitivity limit in the sample of Santini et al. is quite
high ($\rm SFR \gsim 10^{-2} \Msol\,\yr^{-1}$) relative to the model
predictions at lower masses, making it difficult to study the passive
population at high redshift.

The new SF laws, through their impact on the cold gas contents of
galaxies, strongly affect the passive population and its evolution as
seen in the SFR-$M_{\star}$ plane, in a way that could be used as an
important constraint on the SF model, if observational data could
probe low enough SFRs.  On the other hand, the active SF sequence is
much less affected by the choice of SF law, but instead depends on the
assumption about how quickly gas ejected by SN feedback is
reincorporated into the host halo.

\section{Discussion and conclusions}
\label{discussion} 

Improvements in the quality and quantity of observations of 
the spatially-resolved star formation rates and atomic and
molecular gas contents in local galaxies (e.g. \citealt{Kennicutt07};
\citealt{Bigiel08}) have allowed the development of both
empirical (e.g. \citealt{Blitz06}; \citealt{Leroy08}) and theoretical
star formation (SF) models \citep{Krumholz09}. These models prescribe
a relation between the surface density of star formation $\Sigma_{\rm
SFR}$ and gas $\Sigma_{\rm gas}$, often depending also on
other physical quantities such as the stellar surface density and gas
metallicity. We have revisited the SF recipes for quiescent
galaxy formation in galaxy disks used in the galaxy formation model
{\texttt{GALFORM}} and implement new, parameter-free SF laws with the
aim of isolating observations that could be
used to distinguish between SF models ({see \citealt{Lagos11} 
for the predictions of these models for the atomic and molecular hydrogen 
content of galaxies}).

We test the following SF laws: (i) the Schmidt law,
which has the form $\Sigma_{\rm SFR} \propto \Sigma_{\rm gas}^{N}$
with $N=1.4$ (the KS law) and (ii) its variant, the Kennicutt-Schmidt
law, which includes a SF cut-off at low gas surface densities,
motivated by gravitational stability considerations
(\citealt{Toomre64}; the KS.thresh law); (iii) A SF law of the form
$\Sigma_{\rm SFR} \propto \nu_{\rm SF} \, f_{\rm mol} \,
\Sigma_{\rm gas}$, where $f_{\rm mol}$ depends on the hydrostatic
gas pressure (see Eq.~\ref{fmol}; \citealt{Blitz06}; \citealt{Leroy08}),
and $\nu_{\rm SF}$ is an inverse timescale of SF activity in molecular
gas, assumed to be constant (the BR law); (iv) A variant of the
previous SF law, but assuming $\nu_{\rm SF}=\nu_{\rm SF}(\Sigma_{\rm
gas})$ which depends on the total gas surface density (see
Eq.~\ref{fitD09}; the BR.nonlin law).  (v) The theoretical SF law of
\citet{Krumholz09} in which the ratio of surface densities of
molecular to atomic gas depends on the balance between the
dissociation of molecules due to the interstellar far-UV radiation,
 and their formation on the surface of dust grains (the KMT
law). We apply the new SF
laws to the quiescent SF mode which takes place in
galactic disks (see $\S 2.1$).  

We have applied the new SF laws to two variants of  
\texttt{GALFORM}, those of Baugh et al. (2005; Bau05) and
Bower et al. (2006; Bow06). These models have many differences in
their input physics, including different laws for quiescent SF, and
furthermore their parameters were tuned to reproduce different sets of
observational constraints. 
We have left the parameters for all other physical
processes, including SF in starbursts and feedback from supernovae and
AGN, unchanged when running with the new quiescent SF laws, rather
than retuning these other parameters to try to reproduce the original
observational constraints. This allows us to isolate the impact of invoking
different SF laws. Previous attempts to include similar
SF recipes in semi-analytical models have not focused on how galaxy
properties are affected by changing the SF law alone, but have also
made changes to other model parameters (e.g. \citealt{Dutton09};
\citealt{Cook10}; \citealt{Fu10}).

The choice of quiescent SF law by construction affects the quiescent
SF activity.  In the most extreme case, the cosmic SFR density in the
quiescent mode changes by an order of magnitude when compared with the
original model. However, the total SFR density evolution is
remarkably insensitive to the choice of SF law. In the case of the
Bau05 model the new SF laws increase the SFR in the quiescent mode,
while the contrary happens in the Bow06 model. This is compensated for by 
weaker starburst activity in the Bau05 model and more vigorous
starbursts in the Bow06 model.  This adjustment is particularly effective 
in the case of the Bow06 model, which includes starbursts triggered by 
disk instabilities.  A related consequence is the insensitivity of 
the present-day $b_{\rm J}$-band LF and $g-r$ colour distributions 
and the evolution of the $K$-band LF to the choice of quiescent SF law.

Since the gas is consumed at different rates during quiescent
and starburst SF, the change in the quiescent SF law leads to
large differences in the cold gas content of galaxies. For both models the new SF 
laws predict more galaxies with low cold gas masses, fewer with
intermediate cold gas masses and similar or fewer with high cold
masses. This leads to a lower cosmic mean cold gas density at all redshifts. 
For the Bow06 model, these changes help to improve the agreement 
with the local observations by \citet{Zwaan05}.
We find similar changes in the $M_{\rm H}/L_{\rm B}$ gas-to-light ratios as
functions of luminosity.

Finally, we investigated the distribution of galaxies in the
SFR-$M_{\star}$ plane at different redshifts for both the old and new
SF laws.
In most cases, the models predict two sequences in this
plane, an ``active'' sequence with higher SFRs and a ``passive''
sequence with lower SFRs. The form of the ``active'' sequence, which
is dominated by central galaxies, appears insensitive to the quiescent
SF law. This 
can be understood in terms of a rough balance being set up between
accretion of gas due to cooling in the halo and consumption of gas by
star formation and ejection by supernova feedback (at a rate which is
proportional to the SFR). The cold gas mass in a galaxy therefore
adjusts to achieve this balance (for a given SF law), while the SFR
itself is insensitive to the SF law. The form of the active sequence
is, however, sensitive to the timescale for gas which has been ejected
by supernova feedback to be reincorporated into the halo, particularly 
in low-mass galaxies where the supernova feedback is very effective at
ejecting gas. This reincorporation timescale is relatively short
(around a halo dynamical time) in the Bow06 model, but longer in the
Bau05 model. As a consequence, the active sequence extends down to
very low stellar masses in the Bow06 model and its variants with
different SF laws, but it disappears at low masses in the Bau05 model
and its variants.

{The form of the ``passive'' sequence, where
most satellite galaxies are found, is much more sensitive to the SF
law. Satellite galaxies  
are assumed to not accrete any cold gas, thus they are 
simply using up existing reservoirs of cold gas by star
formation and feedback, and their gas contents and SFRs are thus
sensitive to the rate at which this has happened over their 
history. This depends on the SF law and weakly on 
other parameters, such as SN feedback. 
We propose 
that the direct measurement of the (low) SFR of passive galaxies 
offers a constraint on the SF law and can be directly compared 
with the predictions made in this work.}

We made a qualitative comparison of the predicted distribution of
galaxies in the SFR-$M_{\star}$ plane with observational data. For $z
\approx 0$, we compared with the SDSS DR7 results of 
\citet{Brinchmann04} which show {two sequences,  
of active and passive galaxies, at high stellar masses}.
The Bow06 model variants (particularly those 
with the BR, BR.nonlin or the KMT SF laws) appear in better
agreement with these data than the Bau05 model, after 
accounting for the sensitivity limits. 
Observations at higher redshifts indicate the presence of an active 
SF sequence that
evolves in normalization but not in slope
(e.g. \citealt{Brinchmann04}; \citealt{Daddi07}; \citealt{Elbaz07};
\citealt{Stark09}). The model predictions appear in general agreement
with these observations for both the Bau05 and Bow06 models and their
variants. \citet{Rodighiero10} and \citet{Santini09} 
have found indications of
bimodality at $z\sim 1$ in massive galaxies, consistent with our
predictions for the Bow06 model in the three variants already
mentioned.

Future observations of SFRs, stellar masses and particularly the gas
contents of galaxies at high redshifts will be crucial for testing
whether current ideas about star formation laws applied at earlier
stages in galaxy evolution. Telescopes such as ALMA, JWST and ELTs
will make possible measurements of very low SFRs such as those predicted
here for the passive galaxy population, while observations with ALMA,
ASKAP, MeerKAT and eventually SKA will be critical for measuring the
molecular and atomic gas contents of high-redshift galaxies. In this
way we will be able to confirm or rule out the predictions of the
models presented here.

\section*{Acknowledgements}
We thank Shaun Cole, Mark Swinbank, Martin Stringer, 
Jarle Brinchmann, Dave Gilbank and 
Violeta Gonz\'alez for useful comments and discussions. 
We thank the anonymous referee for helpful remarks 
that enable to improve the presentation of this work. CL  
gratefully acknowledges an STFC Gemini studentship. 
AJB acknowledges the support of the Gordon~\&~Betty Moore Foundation. 
This work was supported by a rolling grant from the STFC.

%* Cites during the paper put by hand
%************************************
%************************************
%----------------------------------------------
\bibliographystyle{mn2e_trunc8}
\bibliography{Lagos}
%---------------------------------------------------------------------
\label{lastpage}

\appendix
\section[]{Numerical integration of the star formation equations}\label{numer:integration}

Here we outline the numerical solution of the
equations that describe the changes in the baryonic content of galaxies,
namely the hot gas, cold gas, stellar mass and the metals in each
component.

\subsection{The instantaneous SFR}

The radial profiles of galactic disks
are not resolved in {\texttt{GALFORM}}
(except in special cases e.g. \citealt{Stringer07}). We assume disks are described 
by an exponential profile (see \citealt{Cole00}),

\begin{equation}
\Sigma_{\rm disk} (\rm R) = \Sigma_{0} e^{-\rm R/r_{\rm eff}},
\label{diskprof}
\end{equation}

\noindent where $\rm r_{\rm eff}$ is the scale length of the disk,
$\Sigma_{0}=M_{\rm disk}/2\pi \rm r_{\rm eff}^2$ is the central
surface density and $M_{\rm disk}$ is the disk mass in the component
traced (i.e. the cold gas mass, $M_{\rm cold}$ for $\Sigma_{\rm gas}$,
or the stellar mass of the disk, $M_{\rm stellar, disk}$ for
$\Sigma_{\rm stellar}$).  Note that $\rm r_{\rm eff}$ in an
exponential disk is related to the half mass radius by $\rm r_{50} =
1.67r_{\rm eff}$.

The instantaneous SFR in a given episode will be the surface integral
of $\Sigma_{\rm SFR}$ over the full disk, except for the case of the
KS.thresh SF law, in which we integrate only over the unstable region,
which corresponds to the solutions of $\Sigma_{\rm gas}=\Sigma_{\rm
crit}$ (see Eq.~\ref{gascut}).  In the case that the integral cannot
be solved analytically (as in the BR, BR.nonlin and KMT SF laws),
we use Romberg integration to compute the global SFR by integrating
$\Sigma_{\rm SFR}(\rm R)$ over the whole disk. This method uses
adaptive, equally-spaced divisions in $R$ to achieve the
required accuracy in each calculation. For simplicity, we do not
distribute the newly formed stars only over the part of the disk
where the SF activity occurs, but instead we assume that gas and stars
are always distributed with the same exponential profile. 
\citet{Stringer07} found this to be a reasonable
approximation for the gas component in their radially resolved
calculations of galactic disk evolution. However, this approximation
could overestimate the radial extent of the stellar component of the
disk in gas dominated galaxies (see \citealt{Stringer07}; we revisit
this point in Appendix C2).

\subsection{The SF equations}

The SF activity in {\texttt{GALFORM}} is regulated by three channels: 
(i) accretion of gas which cools from hot gas haloes onto the disk, 
(ii) SF from the cold gas and, (iii) reheating
and ejection of gas due to SNe feedback. These channels modify the
mass and metallicity of each of the mass components (i.e. stellar
mass, $M_{\star}$, cold gas mass, $M_{\rm cold}$, hot halo gas mass,
$M_{\rm hot}$, and their respective masses in metals, $M_{\star}^Z$,
$M_{\rm cold}^Z$ and $M_{\rm hot}^Z$).  The system of equations
relating these quantities is \citep{Cole00}:

\begin{eqnarray}
\dot  M_{\star{\hphantom{col}}}     &=&  (1-R) \psi  \label{eqn:sff} \\
\dot  M_{\rm cold}^{\hphantom{Z}}  &=& \crate - (1-R+\beta) \psi \\
\dot  M_{\rm hot{\hphantom{l}}}   &=& - \crate + \beta \psi \\
\dot  M_{\star{\hphantom{olj}}}^Z   &=& (1-R)  Z_{\rm cold} \psi \\
\dot M_{\rm cold}^Z &=& \crate Z_{\rm hot} \nonumber \\
 &+& (p - (1+\beta-R)Z_{\rm cold}) \psi \\
\dot  M_{\rm hot{\hphantom{l}}}^Z &=& -\crate Z_{\rm hot} +
                       (p\,e + \beta Z_{\rm cold}) \psi .
\label{eqn:sflc}
\end{eqnarray}

\noindent Here $p$ denotes the yield (the fraction of mass converted
into stars that is returned to the ISM in the form of metals), $e$ is the fraction 
of newly produced metals ejected directly from the stellar disk 
to the hot gas phase, $R$ is
the fraction of mass recycled to the ISM (in the form
of stellar winds and SN explosions), $Z_{\rm cold}=M_{\rm
cold}^Z/M_{\rm cold}^{\hphantom{Z}}$ is the metallicity of the cold
gas, and $\beta$ is the efficiency of stellar feedback. Descriptions
the feedback mechanisms can be found in \citet{Benson03}
and \citet{Lacey08}.  These equations are defined as a function of
the instantaneous SFR, $\psi$, that in the standard version of
{\texttt{GALFORM}} has the form given by Eq.~\ref{SFlawGALFORM}, with
$\psi \propto M_{\rm cold}$.  This form of the SFR has the advantage
that Eqs.~$15-20$ can be solved analytically assuming that $\crate$ 
is constant over a timestep.

The new SF laws are non-linear in $M_{\rm cold}$, so they require 
Eqs.~(A2)-(A7) to be solved numerically.  Once 
$\psi$ ($\S 2.3.1$) is calculated, we proceed to integrate Eqs.~(A2)-(A7).  
In {\texttt{GALFORM}}, the equations tracking the
evolution of the baryons are integrated over ``halo timesteps'' given 
by the time resolution at which the halo merger tree is
stored, which in the case of both N-body and Monte Carlo trees is 
independent of the individual SF timescales in each galaxy.  Hence, 
for some cases, the halo timestep could be large compared
to the SF timescale. This makes it necessary to solve Eqs.~(A2)-(A7) 
using adaptive stepsizes to achieve accurate solutions.  We
use the fourth-order Runge-Kutta (RK) method with adaptive stepsizes
\citep{NR}.  The quantities involved in the calculation of the
instantaneous SFR, such as stellar mass, gas mass, metallicity of the
gas, are updated in each sub-step. However, for
simplicity, we assume that the disk scale length $\rm r_{\rm eff}$ and 
$\dot M_{\rm cool}$ (see
Eq.~\ref{diskprof}), remain constant during the integration over each
halo step.  After integrating Eqs.~$15-20$, we infer the
galaxy luminosity by integrating the SF history over a halo timestep,
interpolating the SFR between the values output at the substeps of the RK
integration.

\section[]{The star formation laws}\label{App:sfls}

In this appendix we explain more extensively some physical details 
of the new SF
laws included in the \texttt{GALFORM} model.

\subsection{The critical surface density of the K98 star formation law}\label{App:K98}

In a thin isothermal disk, the critical surface density 
for gravitational instability of axisymmetric perturbations, $\Sigma_{\rm crit}$, 
is given by 

\begin{equation}
\Sigma_{\rm crit}=\frac{\kappa\, \sigma_{\rm g}}{Q_{\rm
crit}\, \pi G},
\end{equation}

\noindent where $\rm Q_{\rm crit}$ is a dimensionless constant $\sim 1$ \citep{Toomre64},
$\sigma_{\rm g}$ is the velocity dispersion of the gas and $\kappa$ is
the epicyclic frequency of the disk.  For realistic gas/stellar disks
we expect $\rm Q_{\rm crit}> 1$ due to the effects of non-axisymmetric
instabilities and the gravity of the stars (see references in K89).
It has been shown that $\sigma_{\rm g}$ is approximately constant
within disks.
This means that the radial dependence of  
$\Sigma_{\rm crit}$ is determined by that of $\kappa$, and can be
estimated directly from a galaxy rotation curve,

\begin{equation}
\kappa=\sqrt{2} \frac{V}{R} \left(1+\frac{R}{V}\frac{{\rm d}V}{{\rm d}R}\right)^{1/2}.
\end{equation}

For a flat rotation curve,

\begin{equation}
\Sigma_{\rm crit}= \frac{\sqrt{2}}{Q_{\rm crit} \,\pi G} \, \sigma_{\rm g}\, \frac{V}{R},
\label{gascut2}
\end{equation}

K89 compared $\Sigma_{\rm crit}$ with the gas profile of the disk
($\Sigma_{\rm gas}$), and found that at the radius of the outermost
HII regions (which indicate recent SF activity), $\Sigma_{\rm
gas}/\Sigma_{\rm crit}\approx 1.9-3.3$ for $Q_{\rm crit}=1$ (after
scaling by our choice of $\sigma_{\rm g}$). The median corresponds to 
$Q_{\rm crit}\approx 2.5$.

\subsection{The midplane hydrostatic pressure of disk galaxies}\label{App:BR}

Under the assumptions of local isothermal stellar and gas layers, and
$\sigma_{\star}>\sigma_{\rm gas}$, the midplane hydrostatic pressure in disks, 
$P_{\rm ext}$, can be approximated to within
$10$\% by \citep{Elmegreen93} 

\begin{equation}
P_{\rm ext}\approx \frac{\pi}{2} G \Sigma_{\rm gas} \left[ \Sigma_{\rm gas} +
\left( \frac{\sigma_{\rm g}}{\sigma_{\star}}\right)\Sigma_{\star}\right], 
\end{equation}
where $\Sigma_{\rm gas}$ and $\Sigma_{\star}$ are
the total surface densities of gas and stars, respectively, and
$\sigma_{\rm g}$ and $\sigma_{\star}$ give the vertical velocity
dispersion of the gas and stars.  We assume a constant gas velocity
dispersion, $\sigma_{\rm g}=10\, \rm km\,\rm s^{-1}$ (see $\S 2.2.1$).
By assuming that $\Sigma_{\star}\gg\Sigma_{\rm gas}$, 
$\sigma_{\star}=\sqrt{\pi \rm G\, \rm h_{\star}\Sigma_{\star}}$, 
where $h_{\star}$ is the stellar scale height.
This
approximation could break down for very high redshift galaxies, whose
disks are gas dominated. In such cases, we assume a floor of
$\sigma_{\star}\ge\sigma_{\rm g}$.  We estimate the stellar
scaleheight assuming that it is proportional to the radial scalelength
of the disk, as observed in local spiral galaxies, with $\rm r_{\rm
eff}/h_{\star} \approx 7.3 \pm 1.2$ \citep{Kregel02}.

\subsection{The KMT star formation law}\label{App:KMT}

KMT theoretical model attempts to deal with three
problems: (i) the determination of the fraction of gas in the 
molecular phase; (ii) the estimation of the characteristic
properties (masses and surface densities) of GMCs, using a mixture of
theoretical ideas and observed correlations; and (iii) the estimation of the
rate at which molecular clouds convert themselves into stars.  The
latter rate is known observationally to be very small ($\approx 1$\%
of mass per free-fall time; \citealt{Krumholz07}) and is 
understood as a result of the regulation of the SF activity by
supersonic turbulence \citep{Krumholz05}.  KMT09 write the SFR per
unit total gas mass as a function of two factors, the fraction of gas
in the molecular phase, $f_{\rm mol}$ (see previous subsection), and the 
SFR per unit molecular mass, $\nu_{\rm SF}$, as in Eq.~\ref{SFlawBR}.

\subsubsection{The molecular-to-total gas ratio}\label{App:KMT1}

The fraction $f_{\rm mol}$ depends on metallicity and the 
gas surface density of the atomic-molecular complex.  
The KMT09 expression for the molecular-to-total gas fraction predicts that
in very metal poor environments, $f_{\rm mol} \to 0$. If implemented 
at face value, this would prevent any SF from taking place in pristine 
gas at very high redshift, and consequently there would be no 
enrichment to make possible any subsequent SF. However, this behaviour 
results from neglecting gas-phase reactions for the formation of $H_2$ 
molecules, which dominate over formation on dust grains when the  
metallicity is very low. Guided by the results of calculations of the 
formation of the first stars (see \citealt{Bromm04} for a review), we 
will therefore assume that a minimum $f_{\rm mol}^{\rm min} = 
10^{-4}$ applies at the onset of the SF activity in our implementation. 
 
\subsubsection{The star formation timescale}\label{App:KMT2}

KMT define the inverse of the timescale needed to consume 
the gas in a cloud into stars as $\nu_{\rm SF} =
\epsilon_{ff}/t_{ff}$, where $t_{ff}$ is the freefall time of a GMC 
and $\epsilon_{ff}$ is the fraction of gas converted to stars per 
freefall time (see \citealt{Krumholz05}). In this model, $\epsilon_{ff}$ is 
a weak function of cloud properties, but the freefall time
$t_{ff}$ depends on the cloud mass $M_{cl}$ and surface density
$\Sigma_{cl}$. 
KMT assume that the GMC mass corresponds to the 
critical Jeans mass for gravitational instability, and that 
GMCs are embedded in a 
gaseous disk that is marginally stable by the \citet{Toomre64} condition
($Q=1$). KMT09 then use an observed correlation found for nearby
galaxies between the global angular velocity of rotation around a
galaxy and the gas surface density, which leads to $M_{cl} \propto
\Sigma_{\rm gas}$.

\section[]{An Illustration of the impact of applying different star
  formation laws}\label{App:SFR}

\begin{figure*}
\begin{center}
\includegraphics[width=0.48\textwidth]{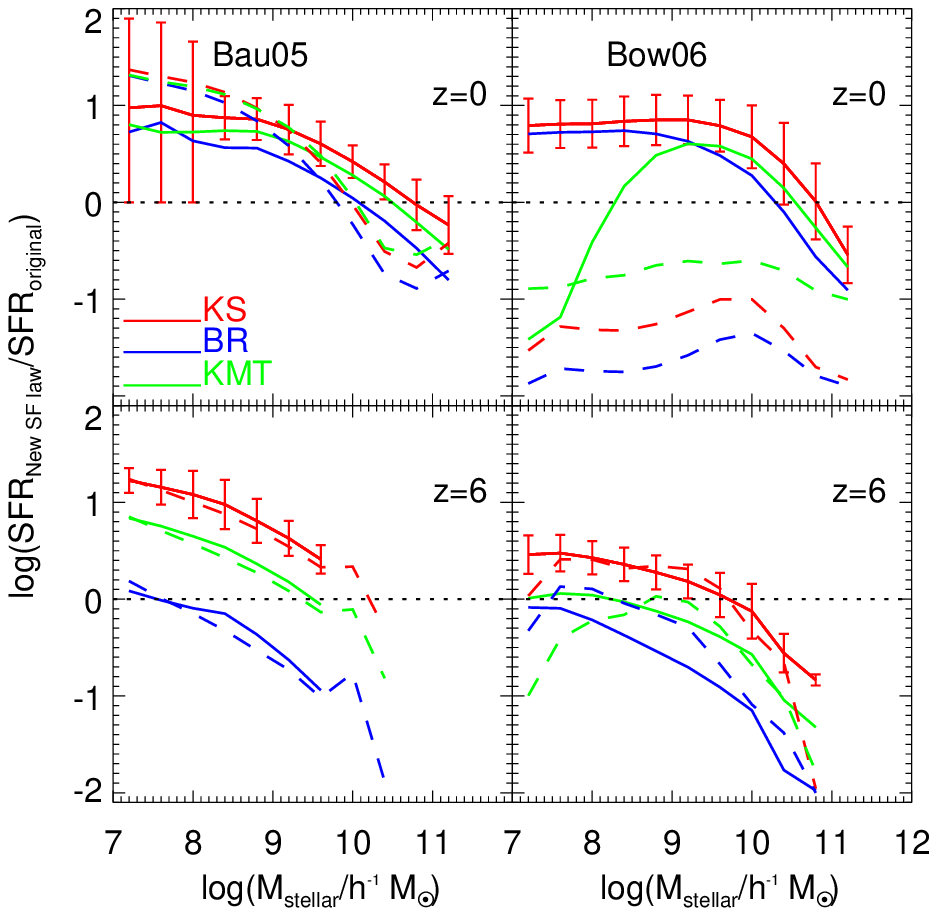}
\includegraphics[width=0.48\textwidth]{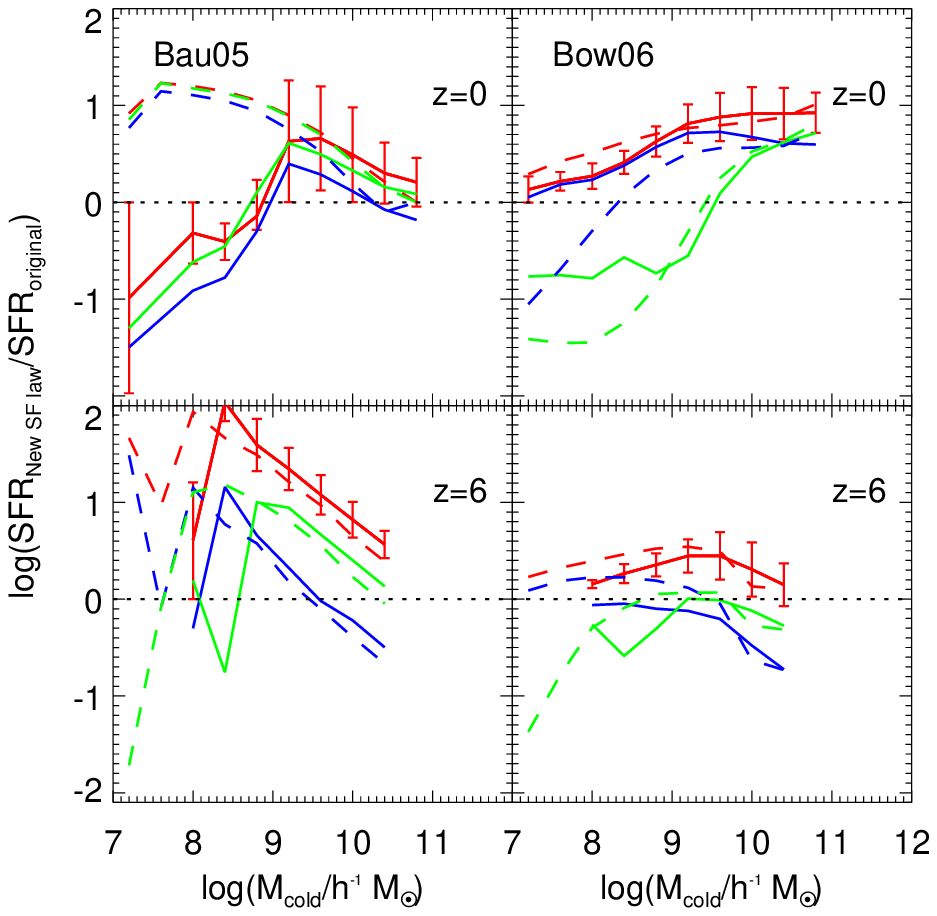}
\caption{Ratio between the new SFR and that in the
original model, $SFR_{\rm New}/SFR_{\rm original}$, plotted as
a function of stellar mass (left-hand four panels) and of cold gas
mass (right-hand four panels).  The upper row shows the predictions at
$z=0$ and the lower row shows $z=6$.  Within each set of
four panels, the left column shows the Bau05 model and the right
column shows the Bow06 model. Three representative SF laws are shown,
the KS law (red), BR (blue) and KMT (green). Solid lines show the
results for central galaxies only, while dashed lines show the results
for satellite galaxies only. Only galaxies with $M_{\rm cold}>0$ are
included in this plot. Error bars show the $10$ and $90$ percentiles
of the distribution, and for clarity are shown only for the KS law.  }
\label{taus}
\end{center}
\end{figure*}

The semi-analytical model follows a range of
physical processes, as set out in $\S 2$. The interplay between
these makes it difficult to disentangle the impact of one
ingredient in isolation. To gain an understanding of the consequences
of changing the SF law, we take the outputs of the original Bau05 and
Bow06 models, and calculate the SFR using the SF
prescriptions in \S\ref{modelssec}, and compare with
the SFR in the original model.  To do this, we freeze the physical
properties of the galaxy used in the SFR calculation (i.e. galaxy
size, stellar mass in the bulge and disk, cold gas mass, rate of
accretion onto the disk of newly cooled gas, and metallicity of the
cold gas) to make a consistent comparison, in which any difference
will be due exclusively to the new SF law.
Fig.~\ref{taus} shows the ratio between the SFR calculated using a
given SF law and that from the original recipe, 
$\rm SFR_{\rm New}/\rm SFR_{\rm original}$, as a function 
of stellar mass (left panel) and cold gas mass (right panel), at $z=0$ and
$z=6$, for the Bau05 and Bow06 models. 

{\it $\rm SFR_{\rm New}/\rm SFR_{\rm original}$ vs. stellar mass}. 
At $z=0$, the new SF laws result
in larger SFR for low stellar mass central galaxies by around an
order of magnitude. The exception is the KMT law in the case of the
Bow06 model, which predicts a lower SFR than in the original model for
low stellar mass galaxies, before rising and peaking around $M_{\rm
stellar}\approx 10^{10} h^{-1} M_{\odot}$. {This arises from the 
abrupt fall in the radial $\Sigma_{\rm SFR}$ predicted by the KMT law due 
to the large disk
sizes of low mass galaxies in the Bow06 model (see Appendix~D2)}.
Moving to high stellar masses, in all cases
the SFRs from the new laws are smaller than the original ones.
In the Bau05 model, similar trends are seen for satellite
galaxies. However, the Bow06 model the predictions for satellites are
quite different. This is because in the Bow06 model,
satellite galaxies have, in general, more modest cold gas reservoirs
compared with the Bau05 model. 
At $z=6$, the predicted 
changes in the SFR are qualitatively similar to those seen at $z=0$. 

{\it $\rm SFR_{\rm New}/\rm SFR_{\rm original}$ vs. cold gas mass}.
The largest differences between the new
and old SF laws are observed at intermediate cold gas masses, 
$8< \log(M_{\rm cold}/ \rm h^{-1}\, \rm M_{\odot})<10$, while the
differences become generally smaller when moving to the extremes of high and low
cold gas masses. This behaviour is more pronounced in the Bau05 model
at all redshifts than in the Bow06 model. 
{The non-linear dependence of the SF laws 
on cold gas mass is directly linked 
with the non-linearity of the SF timescale with cold gas mass. Thus, we expect 
that changing the SF law also affects galaxies 
in different ways depending on their cold gas content.}

\section[]{Other observed properties of galaxies}
\label{localprops}

We compare the predictions of the models with the new SF laws against
selected observations.

\subsection{The galaxy luminosity function}

\begin{figure}
\begin{center}
\includegraphics[width=0.5\textwidth]{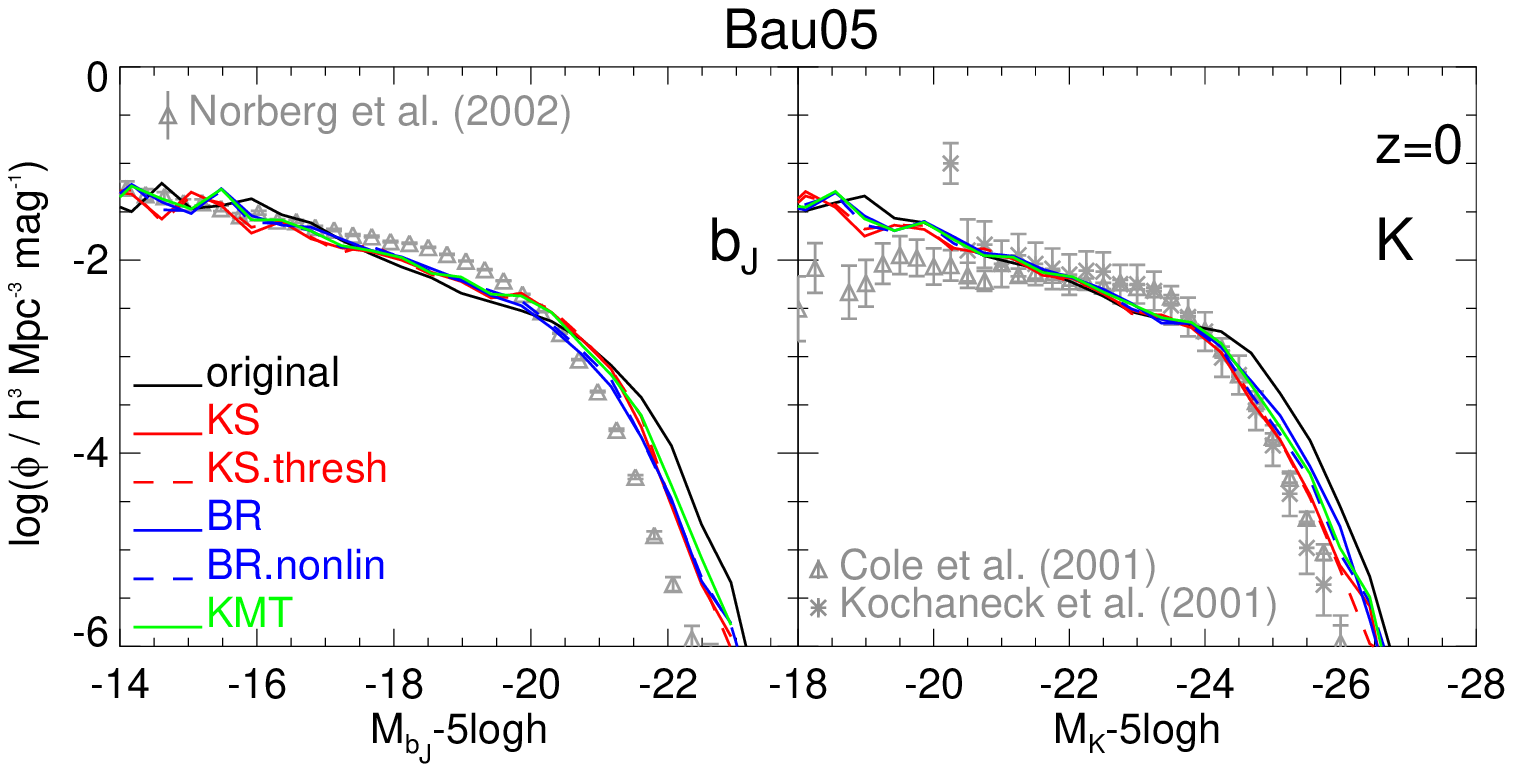}
\includegraphics[width=0.5\textwidth]{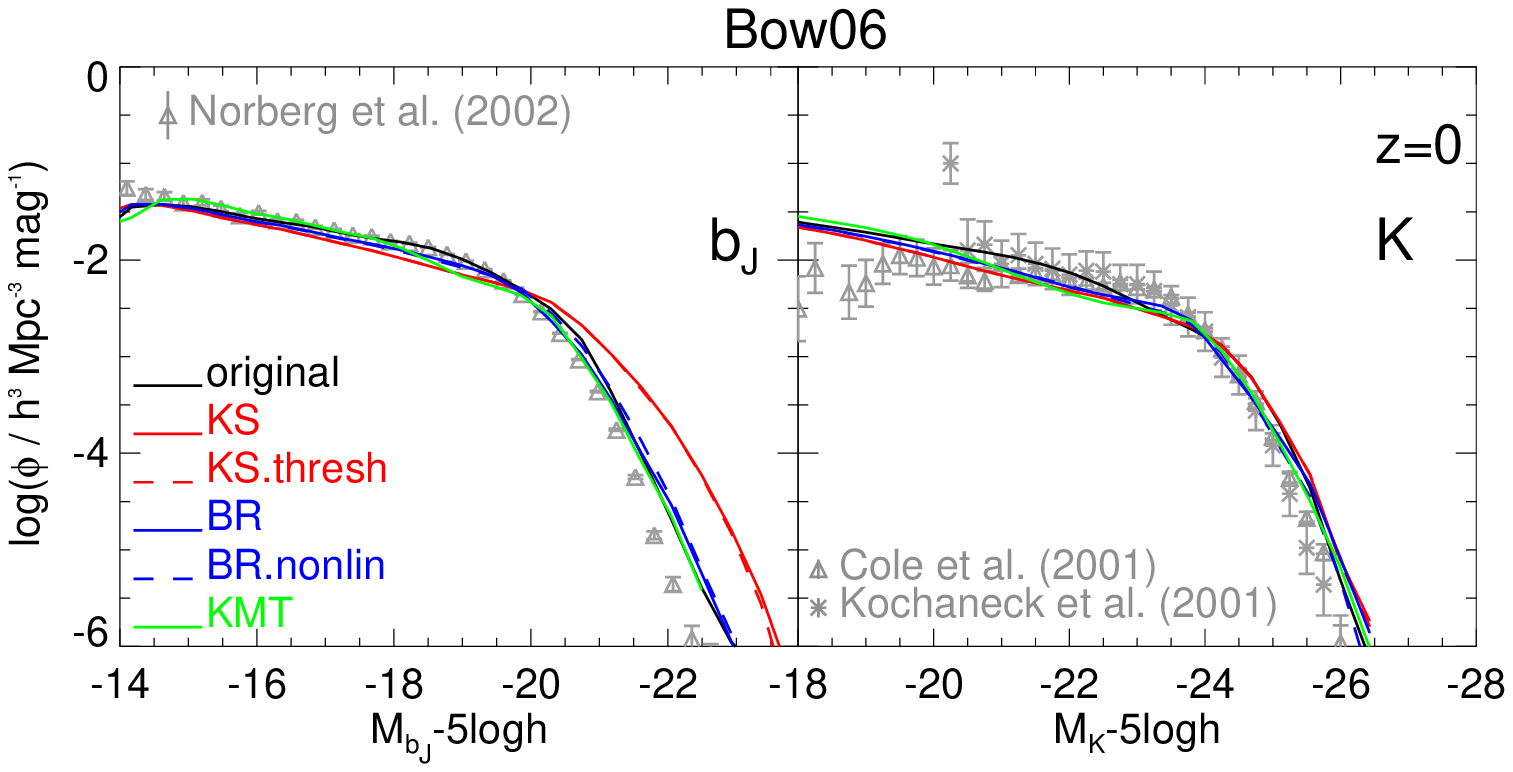}
\caption{The $b_{\rm J}$- (left) and $K$-band (right)
galaxy luminosity functions at $z=0$. The fiducial Bau05 (top)
and Bow06 (bottom) models are shown as black lines. Predictions 
are shown for the KS law (solid red), KS.thresh law (dashed red; 
K98), BR law (solid blue; \citet{Blitz06}), BR.nonlin law 
(dashed blue; \citet{Dutton09}) and the KMT law (solid green; 
\citep{Krumholz09}).  Observational results from \citet{Norberg02} 
($b_{\rm J}$-band) and \citet{Cole01} and \citet{Kochanek01} 
($K$-band) are shown using grey symbols.}
\label{LF}
\end{center}
\end{figure}

The galaxy LFs in the $b_{\rm J}$- and $K$-bands at $z=0$ are shown in
Fig.~\ref{LF}, for the Bau05 (top) and the Bow06 (bottom) models, 
and for variants with new SF laws. Perhaps surprisingly, nearly 
all of the new SF laws tested give a reasonably
good fit to the observed LFs and follow closely the predictions of the
original {\texttt{GALFORM}} models, even though we keep other
parameter values fixed. Exceptions arise for the KS and the KS.thresh
laws in the Bow06 model, which produce too many bright galaxies in the
blue band. 
However, the predicted $K$-band LF is consistent with the
observations even in these cases. This implies that the new laws result in too much SF
activity at the present day in galaxies with $M_{\star}\ge
10^{10} \rm h^{-1}\, \rm M_{\odot}$ for the KS SF laws in the Bow06 model.  
This suggests that the AGN feedback in these objects is not strong enough, and 
there is still gas cooling and hence SF activity at low-$z$. This lack of AGN feedback
is a consequence of the reduced burst SF in these models at $z\le 4$
(see Fig.~\ref{SFRevo}), which reduces the mass of the BHs formed in
massive galaxies and therefore the strength of the AGN feedback ($\S
2$). Another indication of the absence of AGN feedback in the
galaxies producing the excess in the $b_{\rm J}$-band, is that they
correspond to late type galaxies with very low bulge-to-total stellar
mass ratios (that host modest mass BHs).
 
The changes in the LF are particularly mild in the Bow06 model 
on applying the BR, BR.nonlin and KMT laws. This similarity is mainly caused 
by self regulation of the SF channels (i.e. starburst and quiescent SF), in 
which reduced quiescent SF activity at high redshift leads to galaxies characterised by
massive disks and high gas fractions which experience more prominent
SB activity in gas-rich mergers and disk instabilities. Thus the
overall SF remains approximately constant. {This is also reflected in the 
mild impact on the optical colours of galaxies, e.g. $g-r$, in which the blue cloud 
in the colour-magnitude diagram appears slightly 
more pronounced when the new SF laws are applied.}

\begin{figure}
\begin{center}
\includegraphics[width=0.47\textwidth]{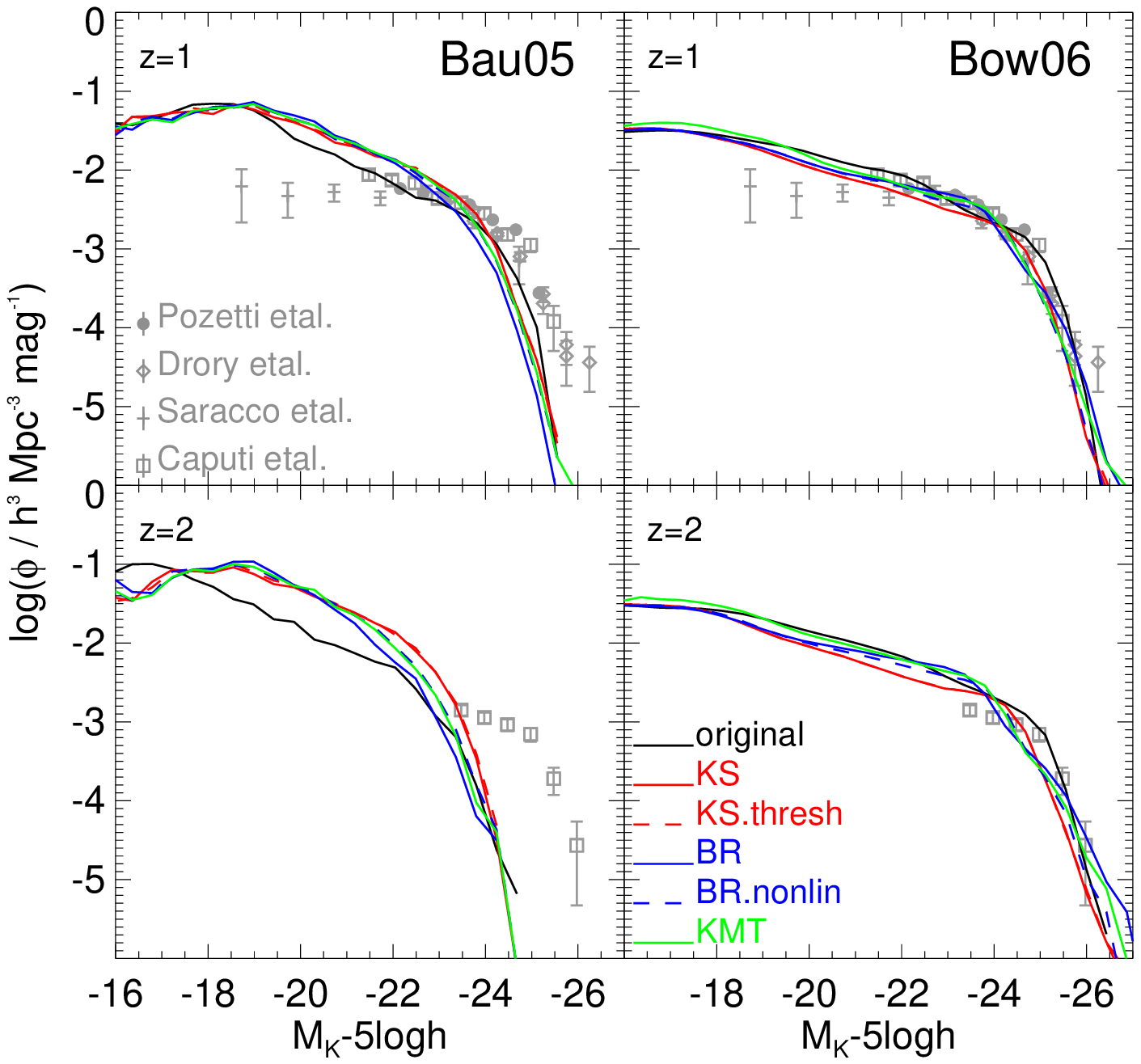}
\caption{ Rest-frame $K$-band galaxy luminosity function at redshifts
$z=1$ and $z=2$ as labelled, for the Bau05 (left) and Bow06 (right) 
models and the different forms of SF laws tested: KS (solid red), KS.thresh (dashed red), BR (solid
blue), BR.nonlin (dashed blue) and KMT (solid green).  Observational
results from \citet{Pozzetti03}, \citet{Drory03}, \citet{Saracco06}
and \citet{Caputi06} are shown as grey symbols, identified by the key
in the top-left panel. }
\label{Kbandevo}
\end{center}
\end{figure}

At high redshifts the similarity in the predicted LFs at $z=0$ is
maintained. Fig.~\ref{Kbandevo} shows the rest-frame $K$-band LF at
$z=1$ and $z=2$ for the original models and variants.
With the new SF laws the Bau05 model produces more galaxies
at intermediate luminosities (i.e. $-16<M_{\rm K}-5\rm log\rm\,
h<-20$) at $z=2$ compared to the original model. This arises 
because the bulk of the SF activity in galaxies is shifted to 
higher redshifts with the new SF laws
(Fig.~\ref{SFRevo}). 
Note that, even though the
LF remains nearly unchanged, the contributions to the total luminosity 
from the bulge
and disk components of galaxies are modified due to the change in the
quiescent and burst SF modes. The new SF laws tend to produce twice
as many ellipticals as in the original models over the
whole mass range.

\subsection{Galaxy sizes}

\begin{figure}
\begin{center}
\includegraphics[width=0.5\textwidth]{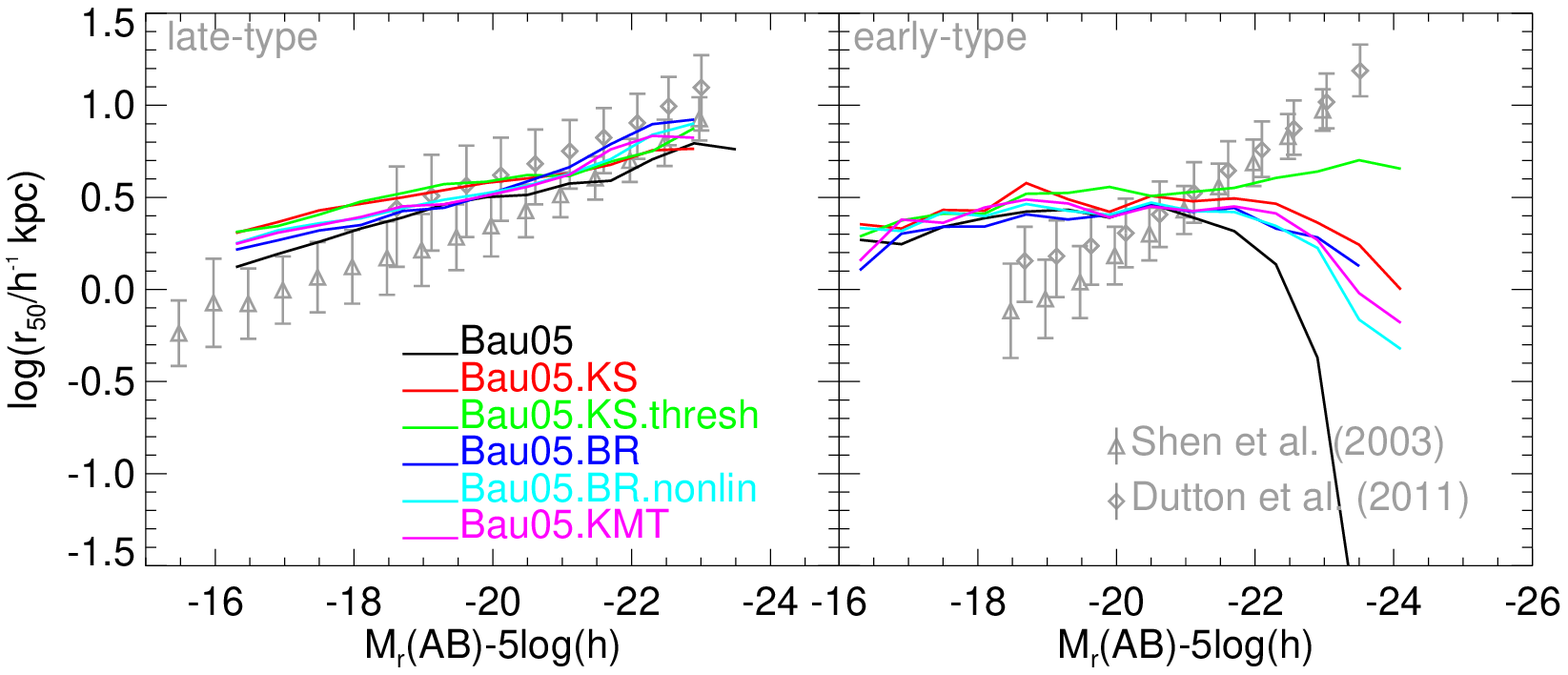}
\includegraphics[width=0.5\textwidth]{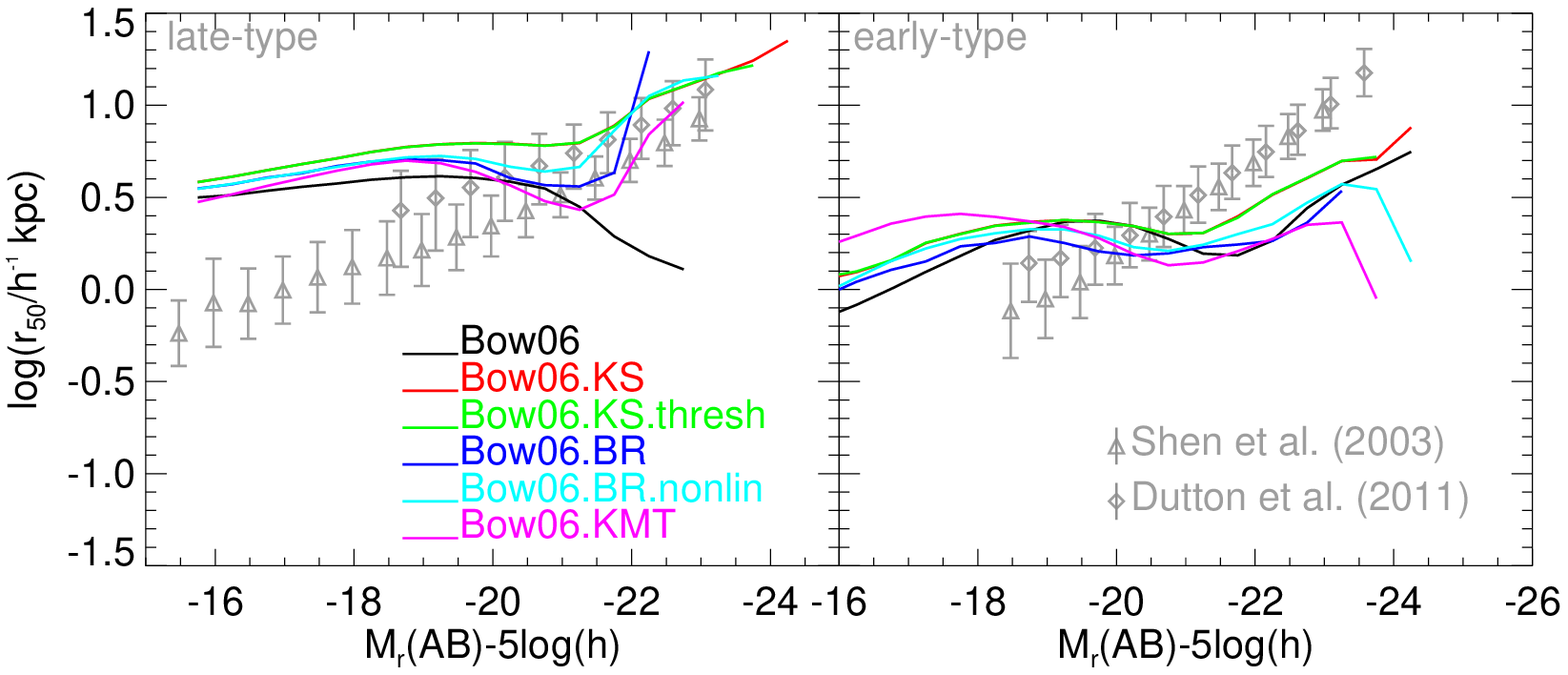}
\caption{Galaxy half light radius as a function of $r-$band magnitude
for late- (top) and early-type (bottom) galaxies. Early-types are
defined to have a bulge-to-total luminosity ratio in the r-band
exceeding $0.5$.  Different lines and colours show the medians in the
models tested as indicated by the key. {The observational results
from \citet{Shen03} and \citet{Dutton11} of late- and early-type
galaxies (classified based on their S\'ersic index and colours,
respectively), which are both based on the SDSS, are shown as grey
symbols with errorbars (corresponding to the medians and $10$\%-$90$\%
of the distributions).}  {The widths of the distributions around
the median in the models are not shown for clarity, but are comparable
to those of the observations (see \citealt{Gonzalez09})}.}
\label{sizes}
\end{center}
\end{figure}

In {\texttt{GALFORM}} the size of a galactic disk is determined by the
conservation of angular momentum of the gas cooling from the halo and
the application of centrifugal equilibrium {in the combined
potential of the disk, bulge and host halo}. Newly cooled gas modifies
the angular momentum of the existing disk.  Fig.~\ref{sizes} shows the
size-luminosity relation of late- (top) and early-type (bottom)
galaxies in the two models (the Bau05 model on the left and Bow06 on
the right), and the variants. Early-type galaxies are defined here as
those with a dust extincted bulge-to-total luminosity ratio in the
$r$-band exceeding $0.5$ ($\rm B/T>0.5$; \citealt{Gonzalez09}) {to
approximately correspond to what is used for the SDSS samples. We also
show two different observational estimates of the size-luminosity
relations \citep{Shen03,Dutton11}, both based on SDSS data.}

{Considering first late-type galaxies, we find that the new SF
laws produce only a small change in sizes for the Bau05 model, but a
large change in the sizes of bright disk-dominated galaxies in the Bow06
model, improving the agreement with observations. At intermediate
luminosities, both models agree somewhat better with the
size-luminosity relation found by \citet{Dutton11} than with that of
\citet{Shen03}.} The larger sizes reported by \citeauthor{Dutton11}
are due to the 2-D fitting they perform to galaxy surface brightness
profiles (i.e.  including separate disk and bulge components, and
inclinations), in contrast with the use of circular apertures to
estimate half-light radii in the case of \citeauthor{Shen03}.

{Bright spirals are larger} in the Bow06 variants because the lower
quiescent SFRs predicted by the new SF laws ($\S 3.2$) result in more
massive disks compared to the original Bow06 model, which then causes
more disk instabilities. (A lower SFR means less SNe feedback and
consequently less cold gas ejected from the disk, hence giving a
larger total mass -- stars plus cold gas -- for the disk). In the
original Bow06 model, bright late-type galaxies typically had quite
large bulge-to-total luminosity ratios, close to the classification
boundary $\rm B/T=0.5$, but the increased incidence of disk
instabilities converts some of these to early-type galaxies. The
late-type galaxies which remain have, on average, larger disk sizes
than before (since smaller disks are more unstable for a given
mass). In the Bau05 model, the differences are smaller since this
model already produces longer SF timescales compared to the Bow06
model, due to the form and parameters used in the SF recipe (see $\S
2.1$). The failure of the Bow06 model to predict small enough sizes
for faint disk-dominated galaxies might be ameliorated with a more
detailed modelling of the radial distributions of stars and gas
(e.g. as in \citealt{Stringer07}).

{For the early-type galaxies, on the other hand, the new SF laws
result in large changes in sizes for bright spheroid-dominated galaxies
in the Bau05 model, but only modest changes in the Bow06 model. The
sizes of spheroids are driven by their formation through mergers and
disk instabilities.}  The flatness of the predicted size-luminosity
relations for spheroid-dominated galaxies in comparison with
observations may result from the simplified calculation of the sizes of
merger remnants, rather than implying that an additional mechanism is
needed (see \citealt{Almeida07}; \citealt{Gonzalez09}).

\end{document}